	   \documentclass[traditabstract]{aa}     
           \usepackage{txfonts}
	   \usepackage{epsf}
           \usepackage{amssymb}
           \newcommand{\ud}{\mathrm{d}}

\begin{document}

   \title{Rotating massive O stars with non-spherical 2D winds}
   
   \author{Patrick~E.~M\"uller\inst{1,2} \and Jorick~S.~Vink\inst{1}}
 
   \institute{Armagh Observatory,
              College Hill,
              Armagh, BT61 9DG,
              Northern Ireland,
              UK\newline
	      \email{pem@arm.ac.uk}
	      \and
	      School of Physical and Geographical Sciences,
              Lennard-Jones Laboratories,
              Keele University,
              Staffordshire,
              ST5 5BG,
	      UK}
   
     \date{Received 11 November 2013 / Accepted 19 February 2014}

\abstract
{We present solutions for the velocity field and mass-loss rates for 2D axisymmetric 
outflows, as well as for the case of mass accretion through the use of the Lambert W-function. 
For the case of a rotating radiation-driven wind the velocity field is obtained analytically using 
a parameterised description of the line acceleration that only depends on radius $r$ at any given 
latitude $\theta$. 
The line acceleration $g(r)$ is obtained from Monte-Carlo multi-line radiative transfer calculations.
The critical/sonic point of our equation of motion varies with latitude $\theta$.
Furthermore, an approximate analytical solution for the supersonic flow of a rotating wind 
is derived, which is found to closely resemble the exact solution. 
For the simultaneous solution of the mass-loss rate and velocity field, we use the iterative method 
of our 1D method extended to the non-spherical 2D case.
We apply the new theoretical expressions with our iterative method to the stellar wind from a 
differentially rotating 40\,$M_{\sun}$ O5--V main sequence star as well as to a 60\,$M_{\sun}$ O--giant star, 
and we compare our results to previous studies that are extensions of the Castor et al. (1975, ApJ, 195, 157) CAK 
formalism. Next, we account for the effects of oblateness and gravity darkening. Our numerical results 
predict an equatorial decrease of the mass-loss rate, which would imply that (surface-averaged) 
total mass-loss rates are lower than for the spherical 1D case, in contradiction
to the Maeder \& Meynet (2000, A\&{}A, 361, 159) formalism that is oftentimes employed 
in stellar evolution calculations for rotating massive stars. To clarify
the situation in nature we discuss observational 
tests to constrain the shapes of large-scale 2D stellar winds.
}

  \titlerunning{Radiation-driven winds from rotating massive stars}

  \maketitle

\section{Introduction}

In recent years considerable progress has been made in our theoretical modelling 
of rotating massive stars (Maeder \& Meynet \cite{maeder:meynet:2012}; Langer \cite{langer}) and our basic 
understanding of spherical radiation-driven winds (Puls et al. \cite{puls:vink}).
However, in order to get a grasp on the non-spherical 2D outflows of 
rotating massive stars, involving B[e] supergiants (Zickgraf et al. \cite{zickgraf:wolf}), classical Be stars (Porter \& Rivinius \cite{porter:rivinius}) 
as well as luminous blue variable (LBV) outflows (Groh et al. \cite{groh:hillier}), it is
paramount to combine the intrinsically 2D nature of rotation and mass loss (e.g. Lovekin \cite{lovekin}; Espinosa \& Rieutord \cite{espinosa:rieutord}).
This is not only required for a basic understanding of massive star evolution, but also for 
linking the oftentimes non-spherical supernova (SN) data with their progenitors 
(e.g. Maund et al. \cite{maund:wheeler}, Hoffman et al. \cite{hoffman:leonard}). 

We need to develop 2D wind models in order to obtain a 
physical understanding of how rotation might affect both the strength and latitudinal dependence of 
their outflows. In turn winds may be able to remove significant quantities of angular momentum, potentially 
down to masses as low as 10-15\,$M_{\sun}$ (Vink et al. \cite{vink:brott}). Whether the mass loss originates from the pole or the equator remains
currently unknown. Yet, is of paramount importance for understanding whether rapid rotation is maintained or 
leads to stellar spin loss (Meynet \& Maeder \cite{meynet:maeder}), highly relevant for our understanding of the progenitor evolution of 
long-duration gamma ray bursts (GRBs).

Previous models of the winds from rotating stars have mostly been 1D models of 
the equatorial flow versus the polar flow, although one 2D numerical calculation has been performed 
by Poe (\cite{poe}).
The 1D model of Friend \& Abbott (\cite{friend:abbott}, hereafter FA) 
concerned the influence of stellar rotation on the hydrodynamics of a stellar wind, involving a 
solution of the fluid equations in the equatorial plane,
which included centrifugal forces. They used a form of the radiation force 
after Castor et al.~(\cite{castor:abbott:klein}, hereafter CAK),
but corrected for the finite angular size of the stellar ``disk'' (see also Pauldrach et al. \cite{pauldrach:al}).  

Bjorkman \& Cassinelli (\cite{bjorkm:cassi}, hereafter BC) 
provided an analytical approximation for the axisymmetric 2D supersonic 
solution (i.e. for the velocity field
and density distribution) of a rotating radiation-driven wind, obtained from 
the FA 1D model of the equatorial flow.
In order to find the streamline trajectories, they rotated the FA 
1D solution (in the equatorial plane) up to the initial
co-latitude $\theta_{0}$ of the streamline at the stellar surface, adjusting 
the equatorial rotation velocity of 
the central star $v_{\rm rot}$ by \mbox{$v_{\rm rot}\sin\theta_{0}$}.
The supersonic solutions obtained this way provided the velocity and density 
as a function of $\theta_{0}$ and radius $r$, i.e.~in a non-explicit form: 
given a location $(r,\theta)$, one needs to find $\theta_{0}$ of the streamline 
that passes through that location, by iteratively solving additional equations.
As a result, they explained how rotation can lead to the production
of a dense equatorial disk around, e.g. Be stars, by means of their 
wind-compressed disk (WCD) model.  

Refinements of the BC model have been also made for simulating the density structure
of rotating O-star winds (e.g.~Petrenz \& Puls \cite{petrenz:puls}).
Owocki et al. (\cite{owocki:cranmer}) showed that the inclusion of non-radial line forces leads to a small polarwards $v_{\theta}$
component, which may inhibit disk formation in Be stars. 
Moreover, Maeder (\cite{maeder}) showed that gravity darkening 
as a result of the Von Zeipel (\cite{zeipel}) effect will generally lead to a polar wind. 
It should be noted that the hydrodynamical wind models of Owocki and colleagues that lead to polar winds 
employ an approximated line driving formula from CAK for 1D.    

A generally prolate wind structure was however confirmed by the sectorial 1.5D wind models 
of Pelupessy et al. (\cite{pelupessy:lamers:vink}) that employed 1D detailed Monte Carlo line acceleration computations of Vink et al. (\cite{vink:dekoter:lamers}).
In their models for B[e] supergiants, Pelupessy et al. also
showed that when models are in close proximity to the bi-stability jump it is possible to overcome 
the polar enhancement due to the Von Zeipel effect, and drive equatorial enhancements, as originally suggested by Lamers \& Pauldrach (\cite{lamers:pauldrach})
for Be and B[e] supergiant disk formation. Cure et al. (\cite{cure:rial}) and Madura et al. (\cite{madura:owocki}) derived 1D hydrodynamical models 
for very rapid rotators (above 75\% of the critical rate) finding a slow solution to the classical CAK theory, which may enable 
disk formation in Be stars and B[e] supergiants, when accounting for the wind bi-stability effects of Vink et al. (\cite{vink:dekoter:lamers}). 
However, again, these models employ a simplified treatment of the line acceleration due to the 2 (or 3) parameter power-law approximation 
due to CAK. What is eventually required in order to resolve the intricate problem of stellar rotation with 
mass loss are 3D Monte Carlo radiative transfer calculations 
in combination with a full hydrodynamic solution. 
Most published models have necessarily made significant assumptions with respect to either 
the line-force calculations or the wind hydrodynamics.  

We suggested a new parametrisation of the line acceleration 
(M\"uller \& Vink \cite{mueller:vink}, hereafter Paper I), 
expressing it as a function of radius rather than of the velocity gradient as
in CAK theory. The implementation
of this formalism allowed for local dynamical
consistency as we were able to determine the momentum
transfer at each location in the wind through the use of Monte
Carlo simulations. 
In Muijres et al. (\cite{muijres:al}), we tested our hydrodynamic wind solutions and velocity laws 
by additional explicit numerical integrations of our fluid equation, also 
accounting for a temperature stratification.
These results were in excellent agreement with both our full and our approximated solutions from 
Paper I. We here build on those results, now deriving 
analytical solutions for the 2D case concerning both the velocity and density structure in
an axisymmetric mass outflow (or inflow) scenario. Furthermore, we extend our iterative 
method from Paper I for the simultaneous solution of the mass-loss rate and velocity field to the 2D case of 
a rotating non-spherical stellar wind. 

We obtain the velocity field fully analytically 
without any previous fits to numerical solutions of the fluid equation of FA, if we neglect the polar velocity
$v_{\theta}$ in our model.
We are justified in doing so as long as the stellar rotation speeds are well below the critical value where
disks are formed, that is, for example for O-star winds, where 
non-spherical outflows have only been detected in a small ($\sim20\%$) minority only involving 
``special'' O-type sub-groups (Oe, Onfp) 
from linear spectropolarimetry observations (Harries et al. \cite{harries:howarth}; Vink et al. \cite{vink:davies}).

However, the non-spherical problem for the flow in the equatorial plane
including the case of the outflow at the pole (where $\mbox{$v_{\theta}\equiv{}0$}$ for symmetrical reasons)
are as well fully analytically solved by our improved 2D wind model
without any restrictions to the velocity components and the rotational speed of the central star.
For the specific case of a non-spherical radiation-driven wind, we do not rely on the 
CAK expression for the radiation force, rather we describe the line acceleration as a function of 
stellar radius $g(r,\theta{})$ at a given constant co-latitude $\theta$.
In addition, the critical point of our stellar wind is the sonic point (depending on latitude) and not the CAK critical point. 
The calculation of $g(r,\theta{})$ is performed through Monte Carlo simulations accounting for multi-line transfer, 
and the wind parameters are solved simultaneously -- in an iterative way -- for each latitude of interest. 

The set-up of the paper is as follows.
In Sects.~\ref{sect-basic-hydro-eq}--\ref{sect-line-acc-term}, the hydrodynamic equations for a non-spherical axisymmetric steady flow
are introduced including a derivation of the mathematical description of the radiative line acceleration as a function of
radius for the case of a rotating radiation-driven wind.
The process for obtaining the fully analytical 2D solutions is described and discussed in Sect.~\ref{sect-analyt-sol-Eq-mot}. 
Here, the velocity field for the entire family of solutions is provided in an explicit general expression
from which the solutions for a rotating radiation-driven wind or mass accretion flux (e.g. collapsing protostellar cloud)
follow as unique trans-sonic solutions through the critical point.
Moreover, an approximate analytical solution for the supersonic flow is presented.
In Sect.~\ref{sect-num-meth}, we describe our numerical computation obtaining the radiative acceleration in our stellar wind models. 
Furthermore, our iterative method for the determination of the consistent solution
for the mass-loss rate in case of a spherical wind is being extended and applied to the wind from a rotating star.
In Sect.~\ref{sect-applic-ostar}, we present the application of our models to a differentially rotating stellar wind from a typical  40\,$M_{\sun}$ O5--V-star,
including the effects of oblateness and gravity darkening, and from a rotating 60\,$M_{\sun}$ O--giant star. 
We discuss the results, before we summarise and discuss our findings in Sect.~\ref{sect-disc}.

 \section{Radiation hydrodynamics of rotating and expanding or collapsing systems}  \label{sect-analytical-theory}

     \subsection{The velocity field}  \label{sect-velo-field}

     The velocity field of the differentially rotating system at location \mbox{$\vec{r}=(r,\,\theta,\,\phi)$}
     can generally be described by its spherical components     
      \begin{equation} \label{gen-velocity-field}
       \vec{v}\,(\vec{r}) = v_{\rm r}\,(\vec{r})\,  \vec{e_{\rm r}}\,(\vec{r}) + v_{\rm \phi}\,(\vec{r})\, 
       \vec{e_{\rm \phi}}\,(\vec{r})
                                      + v_{\rm \theta}\,(\vec{r})\,  \vec{e_{\rm \theta}}\,(\vec{r})  
      \end{equation}  
    with the unit vectors $\vec{e_{\rm r}}$, $\vec{e_{\rm \theta}}$, $\vec{e_{\rm \phi}}$, in radial, polar and
    azimuthal direction, respectively, where the following two presentations of the unit vectors
              \begin{equation} \label{e_r:e_phi}
           \vec{e_{\rm r}}\,(\vec{r}) = \left(
                               \begin{array}{c}
                                 \sin\theta \, \cos\phi \\
                                 \sin\theta \, \sin\phi \\
                                 \cos\theta      
                               \end{array}
              \right) \quad \mbox{and}  \quad  
               \vec{e_{\rm \phi}}\,(\vec{r}) = \left(\hspace{1ex}
                               \begin{array}{c}
                                 \hspace{-1.5ex} -\sin\phi \\
                                    \cos\phi \\
                                      0      
                               \end{array}
              \right)
                \end{equation} 
   will be useful.\footnote{We are here not interested in the presentation of the 
   polar vector $\vec{e_{\rm \theta}}$, 
   because the velocity component $v_{\rm \theta}$ will be later set to zero.}

 \subsection{Basic equations of hydrodynamics} \label{sect-basic-hydro-eq}

      Considering the particular chosen system as a non-viscous\footnote{
      \dots as, e.g., most stellar wind models from early-type massive stars adopt (e.g. CAK, FA, BC).
      However, it should be noted that this restricting assumption of an absent friction 
      excludes the transport of angular momentum in a disk,
      which is mostly dominant in the case of a collapsing system with accretion.},
      i.e., ideal fluid, the momentum equation 
          \begin{equation} \label{momentum-eq}
            \rho\, \frac{{\rm D}\,\vec{v}}{{\rm D}\,t} = \vec{f} - \vec{\nabla}\,p
          \end{equation} 
      is valid (see, e.g., Mihalas \& Weibel Mihalas \cite{mihalas:mihalas}), 
      where $D/D\,t$ is the covariant Lagrangean or co-moving time derivative in
      the fluid-frame of a material element and \vec{v} is its velocity, 
      $\vec{f}$ is the total external body force per volume acting on a mass element of fluid,   
      and $\vec{\nabla}\,p$ is the divergence term of a diagonal isotropic stress tensor
      $\vec{\nabla}\cdot{}\tens{T}$, in which $\tens{T}=-p\,\tens{I}$ and $p$ is the hydrostatic pressure. 
 
      One also needs to consider the
      equation of continuity
              \begin{equation} \label{eq-continuity}
                    \frac{\partial\rho}{\partial{}t} + \vec{\nabla}\cdot{}\left( \rho\,\vec{v}\right) = 0 \, ,
                \end{equation} 
     with the covariant divergence $\vec{\nabla}\cdot{}\vec{v}$ of the velocity vector.

    \subsection{Simplifying assumptions} \label{sect-simpl-assump}
 
     Besides the assumption of an inviscid flow and to account for the axisymmetric and stationary problem,
     we make the further following simplifying assumptions to solve the hydrodynamic equations analytically:
     \begin{enumerate}
       \item The stellar wind (flow) is first assumed to be isothermal to derive the analytical expressions for the hydrodynamic solutions.
                       In this case, the equation of state 
                       \begin{equation}  \label{eq-of-state}
                        p = a^{2} \rho
                       \end{equation}
                       is valid, where $a$ is the isothermal speed of sound and
                      $\rho$ is the density of the wind (system).
                      This assumption, however, will be relaxed later by applying our iterative
                      method (described in Sect.~\ref{sect-num-meth}),
                      to compute the mass-loss rates and the parameters in our analytical wind solutions consistently with
                      the radiation field and ionisation/excitation state of the gas of an outflowing stellar model atmosphere in 
                      non local thermodynamic equilibrium, assuming radiative equilibrium, by the use of the {\sc ISA-Wind} code 
                      allowing a temperature stratification.\footnote{
                   The adjustment of our analytical expressions (for the wind solution) to the non-isothermal model wind
                   is here achieved by deriving an approximated analytical solution for the supersonic wind regime, which introduces a terminal velocity
                   $v_{\infty}$ to our initially assumed isothermal wind, and provides a relationship between $v_{\infty}$ and our wind solution parameters 
                   (cf. explanations in Paper I, Sects.~2.5.4. and 2.5.9.).}
                     Additionally, in Muijres et al. (\cite{muijres:al}), we verified our hydrodynamic solutions, 
                     especially for the case of a spherical wind without rotation, by explicit numerical integrations of our fluid equation
                     also accounting for a temperature distribution.
      \item We assume a stationary axisymmetric flow,
                       since we are interested in a rotating star (system) where rotational effects dominate the flow,  
                       i.e. we set
                       \begin{equation}    
                       \frac{\partial}{\partial{}t} \equiv 0, 
		       \quad \frac{\partial}{\partial{}\phi} \equiv 0,
                        \end{equation}
                        and therefore disregard shocks as well. Furthermore, we exclude the presence of clumps.
       \item For symmetrical reasons, the polar component of the flow velocity should be
                       \begin{equation}  \label{v_theta_neglection}
                          v_{\rm \theta} = 0 \, ,
                       \end{equation}
                       as well as the polar and azimuthal force components,
                       \begin{equation} \label{f_theta}
                         f_{\rm\theta} = 0
                       \end{equation}
                       and
                       \begin{equation} \label{f_phi}
                         f_{\rm\phi} = 0 \, ,
                       \end{equation}
                       respectively,
                       in the equatorial plane and at the pole for an axisymmetric flow.
                       At the pole, the hydrodynamic solutions must pass over into those for the spherical case without rotation.
                       Eqs.~(\ref{v_theta_neglection}) to (\ref{f_phi}) allow us then the seperation of the radial motion for any individual latitude.
                       However, for intermediate latitudes, these approximations do generally not hold:
                       for instance, neglecting the meridional velocity there, restricts the application of our model only to those cases where
                       matter exchange between layers of different latitude (and therefore also the occurrance of a friction between) can be neglected.
                       We therefore apply our model, in the case of stellar winds, only to rotating O-stars with rotational speeds 
                       below the critical value where disks are formed.
                       Moreover, the additional involvement of the distortion of a central star due to its rotation (and consequently the effect of gravity darkening as well) in the course of our investigations, 
                       may result in a non-vanishing radiative force $f_{\rm\theta}$ in $\theta$-direction at all mid-latitudes.
                       Therefore, for winds from those rotating O-stars, we employ our formalism and method (in Sect.~\ref{sect-applic-ostar}) exlusively only on the equatorial plane and the pole,
                       to compute the more accurate wind parameters and mass-loss rates there, where the above assumptions (Eqs.~\ref{v_theta_neglection}--\ref{f_phi}) are satisfied best\footnote{
                       Nevertheless, Gayley \& Owocki (\cite{gayley:owocki}) show how even in a wind that is azimuthally symmetric, a net azimuthal line force may result.}.
                       Then, at latitudes between the pole and equator, all corresponding hydrodynamic quantities adopt values which lie in-between these two extreme values (constraints).
        \item In the case of a wind from a luminous early-type star, the wind is primarily driven by
                      continuum plus line radiation forces,
		      where the radial acceleration on a mass element is
                       \begin{equation}  \label{eq-rad-accel}
                       \frac{f_{\rm r}}{\rho} = - \frac{G\,M}{r^{2}}\, \left(1-\Gamma \right) + g_{\rm rad}^{\rm line}
                       \end{equation}
		       with
		       \begin{equation} \label{def-Gamma}
		        \Gamma := \frac{g_{\rm rad}^{\rm cont}}{g}\, ,
		       \end{equation}
		       the force ratio between the radiative acceleration $g_{\rm rad}^{\rm cont}$ due to the continuum opacity
		       (dominated by electron scattering) and the inward acceleration
		       of gravity $g$.
		       $\Gamma$ is supposed to be independent of radius $r$, may however vary with polar angle $\theta$,
		       and $g_{\rm rad}^{\rm line}\,(r, \theta)$ is the outward radiative acceleration due to spectral lines.
		       $M$ is the mass of the central star.
       \end{enumerate}

  \subsection{Simplified hydrodynamic equations}
       
  \subsubsection{Wind density and mass-loss rate}

     If we use the covariant derivative 
     (see Mihalas \& Weibel Mihalas \cite{mihalas:mihalas}), 
     for spherical coordinates and apply assumptions 2 and 3 to the equation
     of continuity~(\ref{eq-continuity}), we find 
              \begin{equation} \label{eq-continuity2} 
                   \frac{\partial}{\partial{}r}\,\left( r^{2}\,\rho\,v_{\rm r} \right) = 0 
                \end{equation} 
   for a two-dimensional axisymmetric and steady flow.
   
   This equation looks the same as that one, we would get for a one-dimensional, spherically symmetric and
    steady flow,  
    however, this expression is quite more general and only fulfilled
    for a given constant polar angle $\theta$. 
    But similar to a spherically symmetric flow, 
    the integration of Eq.~(\ref{eq-continuity2}) yields
          \begin{equation} \label{Def-Mdot(th)}
         4\,\pi\, r^{2}\, \rho\,(r, \theta)\,  v_{\rm r}\,(r, \theta) = \mbox{const.} =: \dot{M}\,(\theta)\, ,    
           \end{equation} 
    in which here \mbox{$\dot{M} (\theta)$} is not the total mass flux through a spherical shell
    surrounding the star, 
    but its mass flux (or mass-loss rate) at co-latitude $\theta$ through the star surface multiplied by
    \mbox{$4\,\pi\, R^{2}$} (cf.~definition in BC)
    that is conserved for each angle $\theta$.
    The value of $\dot{M}\,(\theta)$ can later be determined by the given values of velocity $v_{\rm r}\,(R, \theta)$ and density $\rho\,(R)$
    at the stellar radius R (or at the surface of an inner core).
    
    The total mass-loss rate must then be
            \begin{equation} \label{total-mass-loss}
                \dot{\cal M} \equiv \int_{4\,\pi} \frac{\dot{M}\,(\theta)}{4\,\pi}\, d\Omega \, ,
            \end{equation} 
    integrated over the solid angle $\Omega$.
    Note that in case of a collapsing system (e.g.~protostellar cloud)
    this mass-loss rate is negative, because the inner core gains mass.     

    Finally, by Eq.~(\ref{Def-Mdot(th)}), we obtain the 2D density distribution
           \begin{equation} \label{rho}  
         \rho\,(r, \theta) = \frac{\dot{M}\,(\theta)}{4\,\pi\,r^{2}\, v_{\rm r}\,(r, \theta)}
                                    = \frac{F\,(\theta)}{{\hat r}^{2}\, v_{\rm r}\,({\hat r}, \theta)} 
           \end{equation}   
     at location \mbox{$(r, \theta)$},
     with the defined flux $F=\dot{M}\,(\theta)/4\,\pi\,R^{2}$ through the star's surface at radius $R$
     and the dimensionless radius ${\hat r}=(r/R)$.

   Please note that all formulae derived in this Sect.~\ref{sect-analytical-theory} are expressed in terms of ${\hat r}$
   referring to the radius R, which is (throughout this section) the stellar (i.e. photospheric) radius
   of the central star (or the inner core radius of any central object, respectively). 
   However, all formulae can generally also be applied with respect to the reference radius R to be the
   inner boundary radius $R_{\rm in}$, from where the numerical computations of the (stellar wind) model start.

   \subsubsection{The azimuthal velocity component and the correction for oblateness}  \label{sect-azimuth-comp-oblateness}
    
     By using the correct contravariant components of acceleration $({\rm D}\,v_{i}/{\rm D}\,t)$ in Eq.~(\ref{momentum-eq}), 
     for spherical coordinates, and replacing them by their equivalent physical components
     (see again, e.g., Mihalas \& Weibel Mihalas \cite{mihalas:mihalas}),
     and applying assumptions 1-4, we obtain the simplified 
     r- and $\phi$-component of the momentum equation
      \begin{eqnarray} \label{Eq-of-motion1a}
                 v_{\rm r}\,\frac{\partial}{\partial{}r}\,v_{\rm r} - \frac{v_{\rm \phi}^{2}}{r}  
                    & = & \frac{f_{\rm r}}{\rho} - \frac{a^{2}}{\rho}\,\frac{\partial{}\rho}{\partial{}r}\, ,     \\    
                 v_{\rm r}\,\frac{\partial}{\partial{}r}\,v_{\rm \phi} + \frac{v_{\rm r}\,v_{\rm \phi}}{r} 
                    & = & 0\, ,   \qquad \mbox{or equivalently} \nonumber\\ 
                  \label{Eq-of-ang-momen1} 
                   \frac{v_{\rm r}}{r}\, \frac{\partial}{\partial{}r}\,\left( r\, v_{\rm \phi}\right) 
                    & = & 0\, ,    
      \end{eqnarray}
    with the external radial force per unit mass $f_{\rm r}$, i.e. the radial acceleration on the mass element in Eq.~(\ref{eq-rad-accel}),
    in case of a stellar wind.
     
   The $\phi$-component of the momentum Eq.~(\ref{Eq-of-ang-momen1}) is nothing more than
   the conservation of angular momentum per unit mass
         \begin{displaymath}
          r\, v_{\phi}\,(r,\,\theta) = \mbox{const.} \, ,   
         \end{displaymath} 
   as one would expect for external central forces and axisymmetry, what we have, of course, supposed before. 
    Then, the last differential equation~(\ref{Eq-of-ang-momen1}) can be solved 
    (i.e.~integrated) immediately and separately from Eq.~(\ref{Eq-of-motion1a}),
    to obtain the unknown velocity component $v_{\phi}$
    \begin{displaymath}
       r\, v_{\phi}\,(r,\,\theta) \stackrel{!}{=} R \, v_{\phi}\,(R,\,\theta) 
    \end{displaymath}
    by choosing an adequate boundary (initial) condition
    \begin{equation}
       v_{\rm rot}\,(\theta{}) := v_{\phi}\,(R,\,\theta)\, ,
    \end{equation}
    i.e. rotational speed of the star (inner core) surface at co-latitude $\theta$.

    Hence, the $\phi$--component of the velocity of a particle at distance ${\hat r}$ in its orbit,
    originating from the stellar surface at co-latitude $\theta$ and ejected with $v_{\rm rot}\,(\theta{})$,
    remaining on the cone surface of constant angle $\theta$, becomes
          \begin{equation}  \label{v_phi}  
         v_{\phi}\,({\hat r}, \theta) = \frac{1}{\hat r} \, v_{\rm rot}\,(\theta{}) \, .
           \end{equation} 
    
    Assuming that the central star surface (or inner core surface)
    behaves like a rotating rigid sphere ($R=$constant), the rotational velocity 
    at co-latitude $\theta$ would then be described by       
     \begin{equation}    
          v_{\rm rot}\,(\theta{}) = V_{\rm rot}\,\sin\theta
     \end{equation}  
    with the equatorial rotation speed $V_{\rm rot}$.
    
    However, due to rapid rotation, the central star (core) can become oblate from the centrifugal forces
    and can then be described as a rotating rigid ellipsoid ($R=R(\theta{})$) with rotational velocity
     \begin{equation}    
          v_{\rm rot}\,(\theta{}) = \frac{R(\theta{})}{R_{\rm eq}} \, V_{\rm rot}\,\sin\theta
     \end{equation}  
    at co-latitude $\theta$,
    where $R_{\rm eq}$ is the radius \mbox{$R(\theta{}=\pi{}/2)$} at the equator.
    
    Here, the stellar (core) radius $R(\theta{})$, depending on latitude and rotation speed 
    is given by (Cranmer \& Owocki \cite{cranmer:owocki}, hereafter CO; Petrenz \& Puls \cite{petrenz:puls}, hereafter PP)
     \begin{equation} \label{R-of-theta}
     R(\theta{}) = \frac{3 R_{\rm p}}{\Omega \sin\theta}
	       \cos\left( \frac{\pi + \arccos\left( \Omega \sin\theta \right)}{3} \right)        
     \end{equation}  
    where $R_{\rm p}$ is the polar radius and assumed to be independent of the rotational velocity, i.e. used as stellar input parameter,
    and $\Omega$ is the normalised stellar angular velocity and defined by
    \begin{equation} \label{angular-vel-eq}
      \Omega = \frac{\omega}{\omega_{\rm crit}} = \frac{1}{\omega_{\rm crit}} \frac{V_{\rm rot}}{R_{\rm eq}}
    \end{equation}
    with the critical angular velocity
    \begin{equation}
     \omega_{\rm crit} = \sqrt{\frac{8\, G M \left( 1 - \Gamma \right)}{27\, R_{\rm p}^{3}}}
    \end{equation}
    and equatorial radius
    \begin{equation}
     R_{\rm eq} = \frac{R_{\rm p}}{1 - V_{\rm rot}^{2} R_{\rm p} / \left( 2\, G M \left( 1 - \Gamma \right) \right)} \, .  
    \end{equation}
    Note that the above definition of $\omega_{\rm crit}$ differs from the break-up velocity 
    $\omega_{\rm crit, spherical}= v_{\rm crit}/R_{\rm p}$ introduced in the following Sect.~\ref{sect-eq-of-mo},
    where the rotational distortion of the stellar surface is neglected.

    \subsubsection{The effect of gravity darkening}  \label{sec-grav-dark}
    
    If one considers the distortion of the central star due to its rotation, one also has to account for
    the effect of gravity darkening caused by its oblateness.
    The early work of von Zeipel (\cite{zeipel}) for distorted stars states that the radiative flux $F\,(\theta{})$
    emerging from the surface at co-latitude $\theta$ is proportional to the local effective gravity 
    \begin{equation} \label{von-Zeipel-eq}
    F\,(\theta{}) \propto g_{\perp}\,(\theta{}) \, .
     \end{equation}
    Therein the normal component of gravity is (see Collins \cite{collins}; or PP)
    \begin{eqnarray} \label{g-perp-eq}
    g_{\perp}\,(\Omega(V_{\rm rot}),\theta) & = & \frac{G M}{R_{\rm p}^{2}}\, \frac{8}{27} 
                                 \left[ \left( \frac{27}{8}\, \left(\frac{R_{\rm p}}{R(\theta)}\right)^{2}
				         - \frac{R(\theta)}{R_{\rm p}}\, \Omega^2\, \sin^{2}\theta \right)^2 \right.  \nonumber \\
					   & & + \left. \Omega^4\, \left( \frac{R(\theta)}{R_{\rm p}}\right)^{2}\, \sin^{2}\theta \,
					        \cos^{2}\theta \right]^{1/2} \, ,
    \end{eqnarray}
    with the stellar radius $R(\theta)$ in Eq.~(\ref{R-of-theta}),
    and the proportional constant
    \begin{equation}
    C\,(\Omega) = \frac{L}{\Sigma\,(\Omega)} 
    \end{equation}
   is given by the constraint that the surface-integrated flux is equal to the total luminosity $L$ of the (oblate) star,
   where  
    \begin{eqnarray}
    \Sigma\,(\Omega) & := & \oint_{A} g_{\perp}\,(\Omega,\theta)\, dA \nonumber \\
                     & \approx & 4 \pi\, G\, M \left( 1 - 0.19696\, \Omega^2 - 0.094292\, \Omega^4 \right.  \nonumber \\
		     & & \qquad \quad \, +  0.33812\, \Omega^6 - 1.3066\, \Omega^8 + 1.8286\, \Omega^{10}  \nonumber \\
                     & & \qquad \quad \, \left. -  0.92714\, \Omega^{12} \right)
    \end{eqnarray}
    is the surface-integrated gravity (see the power series in CO).
    
    Together with the use of the Stefan--Boltzmann law for the flux emitted at co-latitude $\theta$
    \begin{equation}
    F\,(\theta{}) = \sigma_{B}\, T^{4}_{\rm eff}\,(\theta{}) \, ,  
    \end{equation}
    where $\sigma_{B}$ is the Boltzmann constant, we obtain finally the following equation
    for the local effective temperature
    \begin{equation} \label{local-eff-T-eq}
     T_{\rm eff}\,(V_{\rm rot},\theta{}) = \left[ \frac{L}{\sigma_{B}\,\Sigma\,(V_{\rm rot})}\, 
                                           g_{\perp}\,(V_{\rm rot},\theta) \right]^{1/4}  \, .
    \end{equation}
    Through the angular velocity $\Omega$ (cf.~Eq.~(\ref{angular-vel-eq})), 
    the local effective temperature in Eq.~(\ref{local-eff-T-eq}) 
    and the stellar radius $R(\theta)$ in Eq.~(\ref{R-of-theta})
    depends also on the continuum Eddington factor
    $\Gamma$ (cf.~Eq.~(\ref{def-Gamma})) which is, in case of an homogeneous spherical star,
    \begin{equation} \label{Gamma-spherical-star}
     \Gamma = \frac{\sigma_{\rm e}\,L}{4 \pi\, c\, G\, M} \, ,
    \end{equation}
    where $\sigma_{\rm e}$ is the electron scattering cross-section.
    
    Then, this Eq.~(\ref{Gamma-spherical-star}) is also used in our case 
    to calculate a mean value of $\bar{\Gamma}$
    from the prescribed value of $L$ of the non-spherical star,
    to be able to evaluate Eq.~(\ref{R-of-theta}) and Eq.~(\ref{local-eff-T-eq})
    for the stellar parameters
    $R(\theta)$ and $T_{\rm eff}\,(\theta{})$ at a given co-latitude $\theta$ of interest.
    
    However, since the effective gravity and therefore the flux vary over the surface of the rotating star,
    we still need to determine the local value of $\Gamma\,(\theta)$ of the non-spherical star
    that can be defined as
    \begin{equation} \label{local-gamma}
     \Gamma\,(\theta{}) := \frac{\sigma_{\rm e}\,L\,(\theta{})}{4 \pi\, c\, G\, M}
		         = \frac{\sigma_{\rm e}\, \sigma_{\rm B}}{c\, G\, M} \, R^2\,(\theta{})\, T^{4}_{\rm eff}\,(\theta{})
    \end{equation}
    by means of the definition of the latitude-dependent luminosity
    \begin{equation}
     L\,(\theta{}) := 4 \, \pi\, \sigma_{B}\, R^2\,(\theta{})\, T^{4}_{\rm eff}\,(\theta{}) \, ,
    \end{equation}
    which is the luminosity of a corresponding spherical star with radius $R(\theta)$ and effective temperature
    $T_{\rm eff}\,(\theta{})$ of the considered non-spherical star at co-latitude $\theta$.

   \subsubsection{The equation of motion}  \label{sect-eq-of-mo}
    
    Next, we wish to solve the r-component of the momentum Eq.~(\ref{Eq-of-motion1a}),
     i.e.~find an expression for the radial velocity component $v_{\rm r}$
     of the non-spherical axisymmetric steady flow.
     Equation~\ref{Eq-of-motion1a} can then be rewritten
         \begin{equation} \label{Eq-of-mot1-nd}         
          {\hat v_{\rm r}}\,\frac{\partial}{\partial{\hat r}}\,  {\hat v_{\rm r}} - \frac{\hat v_{\rm \phi}^{2}}{\hat r}      
             =  -\frac{{\hat v_{\rm crit}}^{2}}{{\hat r}^{2}} + {\hat g_{\rm rad}^{\rm line}} 
                 -  \frac{1}{\rho}\, \frac{\partial \rho}{\partial{\hat r}}
         \end{equation}
    in non-dimensional form.
    In which the following dimensionless velocities (in units of the isothermal sound speed $a$)
     \begin{equation}   \label{v-crit}       
          {\hat v_{\rm r}} := \frac{v_{\rm r}}{a}\, , \quad {\hat v_{\phi}} := \frac{v_{\phi}}{a}\, , \quad 
          {\hat v_{\rm crit}}\,(\theta{}) := \frac{1}{a}\, \sqrt{\frac{G M}{R\,(\theta)}\,(1 - \bar{\Gamma})}\,   ,
     \end{equation} 
    and dimensionless line acceleration
        \begin{equation}   \label{def-eq-grad-hat} 
           {\hat g_{\rm rad}^{\rm line}}\,(\theta{}) := \frac{R\,(\theta)}{a^2}\, g_{\rm rad}^{\rm line}\,(\theta{})
        \end{equation} 
    are used, where
    $v_{\rm crit}\,(\theta{}=0)$ equals the break-up velocity of the rotating central object (usually without consideration of radiative
    line acceleration terms and rotational distortion).
    By means of Eq.~(\ref{rho}) and applying the chain rule to the function
    $1/v_{\rm r}({\hat r})$, we obtain
     \begin{eqnarray*}    
       \frac{\partial \rho}{\partial{\hat r}}  & = & 
            \left( -\frac{2}{ {\hat r}^{3} }\, \frac{1}{ v_{\rm r}\, ({\hat r}) } - \frac{1}{{\hat r}^{2}}\, \frac{1}{{ v_{\rm r}\,
                ({\hat r})}^{2}} \, \frac{\partial v_{\rm r}\,({\hat r})}{\partial{\hat r}}    \right)\, F \\
         & \equiv & - \rho\,({\hat r},\theta)\, \left( \frac{2}{{\hat r}} 
                                                   + \frac{1}{v_{\rm r}\,({\hat r},\theta)}\, 
                                                   \frac{\partial v_{\rm r}\,({\hat r},\theta)}{\partial{\hat r}}
                                                    \right)  \, .                 
     \end{eqnarray*} 
     Using this expression for \mbox{$\partial \rho / \partial{\hat r}$} in Eq.~(\ref{Eq-of-mot1-nd})
     together with our relation for the azimuthal velocity, Eq.~(\ref{v_phi}),
     we finally find
     the dimensionless differential equation of motion (EOM)
     for the radial velocity at constant co-latitude $\theta$
         \begin{equation}  \label{Eq-of-mot2}  
         \left( {\hat v_{\rm r}} - \frac{1}{ {\hat v_{\rm r}} }\right)\, \frac{\partial}{\partial\,{\hat r}}\, {\hat v_{\rm r}} =
                 \frac{{\hat v_{\rm rot}}^{2}\,(\theta)}{{\hat r}^3}  -\frac{{\hat v_{\rm crit}}^{2}\,(\theta)}{{\hat r}^{2}} + \frac{2}{\hat r} 
                + {\hat g_{\rm rad}^{\rm line}}\,(\theta) \, ,
	\end{equation}
     that is now independent of $\rho$.

   \subsection{The line acceleration term and the final equation of motion}  \label{sect-line-acc-term}

\begin{figure}
\centerline{\hspace{0cm}\epsfxsize=9cm \epsfbox{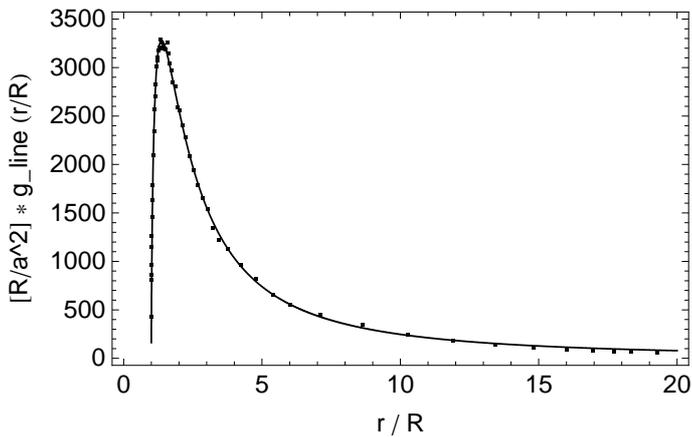}}
\caption{Dimensionless radiative line acceleration ${\hat g_{\rm rad}^{\rm line}}\,({\hat r},\theta{}=\pi/{}2)$ 
         vs. radial distance ${\hat r}$ (in units of $R=11.757\,R_{\sun}$) in the wind from an undistorted O5--V-star
	 rotating with $V_{\rm rot}=500$~km s$^{-1}$ at the equator
	 (see stellar parameters in Table~\ref{table1-par-nodist-star} in Sect.~\ref{sect-applic-ostar}).
	 The dots represent the results from a numerical calculation of ${\hat g_{\rm rad}^{\rm line}}\,({\hat r}_{i},\theta{})$
	 for discrete radial grid points ${\hat r}_{i}$.
	 To determine the line acceleration parameters ${\hat g_{0}}$, $\gamma$, $\delta$ and ${\hat r_{0}}$
	 in Eq.~(\ref{line-acc-term1}), 
	 these values were fit to this non-linear model equation resulting in 
         the displayed fitting curve (solid line):
	 see converged wind parameters in Table~\ref{table1-par-nodist-star} (according to $v_{\infty}=2720$~km s$^{-1}$) 
	 at the end of the iteration process
	 described in Sect.~\ref{sect-it-method} and (lower part of) Table~\ref{tablA1-wind-non-dist}, respectively.
	 }
\label{pic-gline-fit}
\end{figure}
    
    To derive a sophisticated mathematical expression for the radiative line acceleration \mbox{${\hat g_{\rm rad}^{\rm line}}\,({\hat r},\theta)$}
    at a constant co-latitude $\theta$ of interest as a function of radius $r$ only,
    we demand the same physically motivated mathematical properties as described in Paper I (as for 
    the case of a radiation-driven spherical wind):
    \begin{equation} \label{line-acc-term1}
        {\hat g_{\rm rad}^{\rm line}} ({\hat r},\theta) 
         = \frac{\hat g_{0}}{\hat r^{1+\delta}}\, \left( 1 - \frac{\hat r_{0}}{{\hat r}^{\delta}} \right)^{\gamma}
	 \equiv 
	 \frac{\hat g_{0}}{\hat r^{1+\delta\,\left( 1+\gamma \right)}}\, \left({{\hat r}^{\delta}} - {\hat r_{0}}\right)^{\gamma}  \, .
    \end{equation}
   This function is independent of ${\hat v_{\rm r}}$ and \mbox{($\partial\,{\hat v_{\rm r}} / \partial\,{\hat r}$)}
   and dependent on ${\hat r}$ only, at constant co-latitude $\theta$.
   Note that, herein, the set of four parameters
   all depend on latitude: ${\hat g_{0}}={\hat g_{0}}(\theta)$, ${\hat r_{0}}={\hat r_{0}}(\theta)$,  
   $\gamma{}=\gamma{}(\theta)$, $\delta{}=\delta{}(\theta)$, due to the rotation of the central star.
   
   The line force is zero at radius ${\hat r'}={\hat r_{0}}^{1 / \delta}$ near the stellar photosphere (${\hat r'}\approx{}1$)
   and everywhere else positive for ${\hat r}>{\hat r'}$.
   To guarantee the decrease of ${\hat g_{\rm rad}^{\rm line}}\,({\hat r})$ as ${\hat g_{0}}/{\hat r}^{2}$
   with increasing radial distance ${\hat r}$ from the central star
   at intermediate radii (at the right side of the ${\hat g_{\rm rad}^{\rm line}}$ peak) in particular,
   we had to introduce the parameter $\delta$ in addition to $\gamma$
   (where $0<\gamma\lesssim{}1$ and $0<\delta\lesssim{}1$).
   
   Hence, the equation of motion~(\ref{Eq-of-mot2}) for each latitude becomes 
    \begin{equation}  \label{Eq-of-mot3}  
         \left( {\hat v_{\rm r}} - \frac{1}{ {\hat v_{\rm r}} }\right)\, \frac{\partial}{\partial\,{\hat r}}\, {\hat v_{\rm r}} =
               \frac{{\hat v_{\rm rot}}^{2}}{{\hat r}^3}  -\frac{{\hat v_{\rm crit}}^{2}}{{\hat r}^{2}} + \frac{2}{\hat r} 
                + \frac{\hat g_{0}}{\hat r^{1+\delta\,\left( 1+\gamma\right)}}\, 
		\left({\hat r}^{\delta} - {\hat r_{0}}\right)^{\gamma}  \, .
    \end{equation}
   $v_{\rm rot}$ is the azimuthal velocity of the surface of the inner rotating star (object) at co-latitude $\theta$ of interest.
   This equation is fully solvable analytically as in the spherical case without rotation (cf.~Paper I).

    \subsection{Analytical solutions of the equation of motion} \label{sect-analyt-sol-Eq-mot}

    \subsubsection{The critical point and critical solutions} \label{sect-crit-point-sol}

\begin{figure}
\centerline{\hspace{0cm}\epsfxsize=9.5cm \epsffile{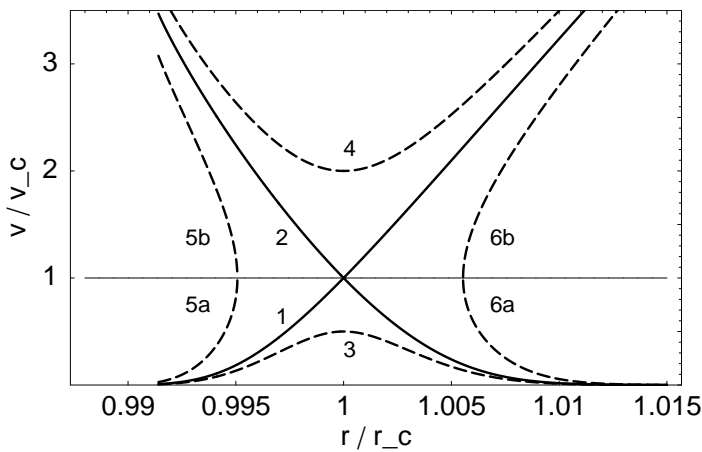}}
\caption{Topology of solutions $|v({\hat r}/{\hat r}_{\rm c},\theta{}=0)/a|$ of the equation of motion, Eq.~(\ref{Eq-of-mot3}),
                vs. radial distance in units of the critical radius ${\hat r}_{\rm c}$ at the \emph{pole}, 
	        for a typical O5--V-star in the centre with
		the line acceleration parameters 
		${\hat g_{0}}=17661$, $\gamma=0.4758$, $\delta{}=0.6878$ and ${\hat r_{0}}=1.0016$ in Eq.~(\ref{line-acc-term1}),
		according to $v_{\infty}=3232$~km/s.
		The horizontal line marks the critical velocity (i.e. sound speed $v_{\rm c}=a$).
                Solution 1 is the unique trans-sonic stellar wind solution through the critical point at ${\hat r}_{\rm c}={\hat r}_{\rm s}=1.0110$
		and ${\hat v}({\hat r}_{\rm c})=1.0$.
                For the description of the different solutions of type 2--6,
                see the discussion in Sects.~\ref{sect-crit-point-sol} and~\ref{sect-sol-eq-mot}.}
\label{sol-topology}
\end{figure}

     The EOM (Eqs.~\ref{Eq-of-mot2} and~\ref{Eq-of-mot3})
     yields several families of solutions that have quite different
     mathematical behaviour and physical significance (cf.~Fig.~\ref{sol-topology}). 
     
     The left hand side of Eq.~(\ref{Eq-of-mot3}) vanishes for $(\partial{\hat v_{\rm r}}/\partial{\hat r}\neq 0)_{{\hat r_{c}}}$
     at the critical radius ${\hat r_{\rm c}}\,(\theta)$, where ${\hat v_{\rm r}}({\hat r}_{c})\equiv{}{\hat v_{\rm r}}({\hat r}_{s})=1$. 
     That is, the critical point velocity here is equal to the isothermal sound speed ${\hat v}=1$,
     and the critical radius is just the sonic radius
     \begin{equation}
         {\hat r_{\rm c}} \equiv {\hat r_{\rm s}} \, ,
     \end{equation}
     as is also the case for thermal winds or mass accretion events (where ${\hat g_{\rm rad}^{\rm line}}\approx{}0$).
     
     We are now interested in under which conditions one can obtain a continuous and smooth trans-sonic flow
     through the critical point ${\hat r}_{\rm c}$ of Eq.~(\ref{Eq-of-mot3}).
     For the case of a stellar wind, this means how to obtain a smooth transition from subsonic and subcritical
     flow \mbox{(${\hat v_{\rm r}}<{\hat v}_{\rm c}=1$)} at small 
     \mbox{${\hat r}<{\hat r}_{\rm c}$}
     to supercritical and supersonic flow \mbox{(${\hat v_{\rm r}}>{\hat v}_{\rm c}$)} at large
     \mbox{${\hat r}>{\hat r}_{\rm c}$}, 
     when this critical solution has a finite positive slope \mbox{($\partial{\hat v_{\rm r}}/\partial{\hat r})>0$}
     at \mbox{${\hat r}={\hat r}_{c}$} (cf.~solid curve 1 in Fig.~\ref{sol-topology})?\footnote{We furthermore assume 
                                                    that both ${\hat v_{\rm r}}$ and \mbox{($\partial{\hat v_{\rm r}}/\partial{\hat r}$)} are
                                                    everywhere single-valued and continuous.}
     Then, it is evident from the left hand side of Eq.~(\ref{Eq-of-mot3}) that one can obtain such a trans-sonic 
     wind,
     if the right hand side (1)~vanishes at the critical radius ${\hat r}_{\rm c}$,
     (2)~is negative for \mbox{${\hat r}<{\hat r}_{\rm c}$}, and 
     (3)~is positive for \mbox{${\hat r}>{\hat r}_{\rm c}$}.  
 
     The opposite situation occurs for the case of mass accretion in e.g. a collapsing cloud.
     If \mbox{($\partial{\hat v_{\rm r}}/\partial{\hat r})_{\hat r_{c}}<0$}, we obtain the second unique trans-sonic and critical solution
     in which ${\hat v_{\rm r}}\,({\hat r})$ is monotonically decreasing from supersonic speeds 
     for \mbox{${\hat r}<{\hat r}_{s}$}, e.g. near the protostar, to subsonic speeds for
     \mbox{${\hat r}>{\hat r}_{s}$} at the outer edge of the cloud (see also the second solid line 2, in Fig.~\ref{sol-topology},
     for the case of a corresponding accretion flow with a star as the central object).
     
     Here we are mainly interested in the critical wind solution
     of Eq.~(\ref{Eq-of-mot3}).
     The right hand side of Eq.~(\ref{Eq-of-mot3}) vanishes at 
     the critical radius ${\hat r}_{c}\,(\theta)$ that solves the equation
     \begin{equation}  \label{crit-rad-eq}   
    2\, {\hat r}^{\delta\,\left(1+\gamma\right) +1} - {\hat v}_{\rm crit}^{2}\, {\hat r}^{\delta\,\left(1+\gamma\right)} 
         + {\hat g_{0}}\,{\hat r}\,\left( {\hat r}^{\delta} - {\hat r_{0}}\right)^{\gamma} + 
	   {\hat v_{\rm rot}}^{2}\, {\hat r}^{\delta\,\left(1+\gamma\right) -1} = 0\, .
     \end{equation} 
   Therefore, the critical radius (for each latitude) has to be determined numerically by means of the above equation and
   the line term pa\-ra\-me\-ters ${\hat g_{0}}$, $\gamma$, $\delta$ and ${\hat r_{0}}$, depending on the rotational speed $V_{\rm rot}$
   of the central object.
   However, if one assumes values of $\gamma$ and $\delta$ close to $1$,
   one can provide a good analytical approximation (for the solution of
   Eq.~(\ref{crit-rad-eq})) for the critical radius
     \begin{equation}  \label{approx-crit-rad}   
      {\hat r}_{\rm c} \approx \frac{1}{4}\,\left( \left( {\hat v}_{\rm crit}^{2} - {\hat g_{0}}\right) 
      + \sqrt{\left( {\hat v}_{\rm crit}^{2} - {\hat g_{0}}\right){}^2
      + 8\,\left( {\hat g_{0}}\, {\hat r_{0}} - {\hat v_{\rm rot}}^{2}\right) } \right)    \, ,
     \end{equation} 
   if the rotation speed at co-latitude $\theta$ fulfills the condition   
     \begin{displaymath}
      {\hat v_{\rm rot}} < \sqrt{ \frac{1}{8}\,\left( {\hat v}_{\rm crit}^{2} - {\hat g_{0}}\right){}^2 + {\hat g_{0}}\, {\hat r_{0}}} \, .
     \end{displaymath}
   For the simpler case of a thermal wind (or any other non line-driven mass flow), 
   where ${\hat g_{\rm rad}^{\rm line}}$ can be set to zero (in Eq.~(\ref{crit-rad-eq})),
   we obtain the analytical solution
     \begin{equation} \label{rc-thermalwind}
      {\hat r}_{\rm c} = \frac{1}{4}\,\left( {\hat v}_{\rm crit}^{2} + \sqrt{{\hat v}_{\rm crit}^{4} - 8\,{\hat v_{\rm rot}}^{2} } \right)\, .
     \end{equation}
   Equation~(\ref{rc-thermalwind}), and Eq.~(\ref{approx-crit-rad}) under the assumption of a constant remaining
   line force (i.e. ${\hat g_{0}}$ value), imply that the critical radius moves closer to the inner core radius
   (at any latitude besides of that of the pole) with increasing rotational speed $V_{\rm rot}$ for these particular 
   cases of a trans-sonic flow.

   \subsubsection{Solving the equation of motion}
   
    The equation of motion~(\ref{Eq-of-mot3}) can be solved by first integrating the left hand side over
    ${\hat v_{\rm r}}$, and then integrating the right hand side over ${\hat r}$, separately,
    which yields   
         \begin{eqnarray} \label{derive-v-sol-1} 
     {\hat v_{\rm r}}^{2} - \ln{\hat v_{\rm r}}^{2} 
              & = & 2\frac{{\hat v}_{\rm crit}^{2}}{\hat r} + 4\, \ln{\hat r}   \nonumber\\
              & & + \frac{2}{\hat r_{0}}\,\frac{\hat g_{0}}{\delta\,\left(1+\gamma\right)} 
	          \left( 1 - \frac{\hat r_{0}}{{\hat r}^{\delta}}\right)^{1 + \gamma} 
		  - \frac{{\hat v}_{\rm rot}^{2}}{{\hat r}^2}  + C \, ,
         \end{eqnarray} 
    with the right hand side of Eq.~(\ref{derive-v-sol-1}) denoted as the function
     \begin{eqnarray} \label{def-f-func}
     f\,({\hat r},\theta{}; {\hat r}', {\hat v_{\rm r}}') \!\!\! &:=& 2\frac{{\hat v}_{\rm crit}^{2}}{\hat r} + 4 \ln{\hat r} \nonumber\\
     & & + \frac{2}{\hat r_{0}} \frac{\hat g_{0}}{\delta\left(1+\gamma\right)} 
           \left( 1 - \frac{\hat r_{0}}{{\hat r}^{\delta}}\right)^{1 + \gamma} \!\!\!\! - \frac{{\hat v}_{\rm rot}^{2}}{{\hat r}^2} 
	 + C\,({\hat r}',\theta{},{\hat v_{\rm r}}')
     \end{eqnarray}
     with the constant of integration $C$, that is determined by the boundary condition of the radial velocity
    ${\hat v_{\rm r}}'$ at a given location $({\hat r}',\theta)$.  
       
     From Eq.~(\ref{derive-v-sol-1}), we can
     determine $C({\hat r}',\theta{},{\hat v_{\rm r}}')$ for the solution that passes through the particular point 
     $({\hat r}',\theta{},{\hat v_{\rm r}}')$,
     \begin{eqnarray}
     C\,({\hat r}',\theta{},{\hat v_{\rm r}}') & = & {\hat v_{\rm r}}'^{2} - \ln{\hat v_{\rm r}}'^{2} - 2\frac{{\hat v}_{\rm crit}^{2}}{\hat r'} 
                                                       - 4\, \ln{\hat r}' \nonumber\\
	                                        & & - \frac{2}{\hat r_{0}}\,\frac{\hat g_{0}}{\delta\left(1+\gamma\right)} 
                                                 \left( 1 - \frac{\hat r_{0}}{{\hat r}'^{\delta}}\right)^{1 + \gamma}
						 + \frac{{\hat v}_{\rm rot}^{2}}{{\hat r}'^2} \, .
      \end{eqnarray}
    Therefore the function $f$ in Eq.~(\ref{def-f-func}) becomes
         \begin{eqnarray}  \label{func-f}
     f\,({\hat r},\theta; {\hat r}',{\hat v_{\rm r}}') & = & {\hat v_{\rm r}}'^{2} - \ln{\hat v_{\rm r}}'^{2} 
                                               + 2\,{\hat v}_{\rm crit}^{2} \left( \frac{1}{\hat r} - \frac{1}{\hat r'}\right)  
					       + 4\,\ln\left(\frac{\hat r}{\hat r'}\right)
                                                \nonumber\\
                                      &  & + \frac{2}{\hat r_{0}}\,\frac{\hat g_{0}}{\delta\left(1+\gamma\right)}  
				             \left[ \left( 1 - \frac{\hat r_{0}}{{\hat r}^{\delta}}\right)^{1 + \gamma}
					                - \left( 1 - \frac{\hat r_{0}}{{\hat r}'^{\delta}}\right)^{1 + \gamma} \right] 
					        \nonumber\\		
				      &  & + {\hat v}_{\rm rot}^{2} \, \left( \frac{1}{{\hat r}'^2} - \frac{1}{{\hat r}^2}\right) 	
					\, .
	\end{eqnarray} 
       And from this, Eq.~(\ref{derive-v-sol-1}) now reads 
     \begin{displaymath}
       {\hat v_{\rm r}}^{2} - \ln{\hat v_{\rm r}}^{2} = f\,({\hat r},\theta; {\hat r}',{\hat v_{\rm r}}')
     \end{displaymath}
     or, equivalently,
       \begin{equation}  \label{derive-v-sol-2}
           -{\hat v_{\rm r}}^{2}\, {\rm e}^{-{\hat v_{\rm r}}^{2}} = - {\rm e}^{- f\,({\hat r},\theta; {\hat r}',{\hat v_{\rm r}}')}\, ,   
       \end{equation} 
     which is solved explicitly and fully analytically in terms of the Lambert W function 
     (cf.~Corless et al. \cite{corless:et_al:93}, \cite{corless:et_al:96}).

    \subsubsection{The solution(s) of the equation of motion}  \label{sect-sol-eq-mot}
    
     It is now possible to provide an explicit analytical expression for the
     solution ${\hat v_{\rm r}}$ of the equation of motion~(\ref{Eq-of-mot3}),
     or Eq.~(\ref{derive-v-sol-2}),
     by means of the W function (cf.~Paper I).
     If we compare Eq.~(\ref{derive-v-sol-2}) with
    the defining equation of the Lambert W function
         \begin{equation}  \label{Defgl-W}
             {\rm W}_{k} (x)\, {\rm e}^{{\rm W}_{k} (x)} = x \, ,
         \end{equation}
    we find that
      \begin{displaymath}
     -{\hat v_{\rm r}}^{2} = {\rm W}_{k} (x)
      \end{displaymath}
    or
    \begin{equation} \label{vr-gensol-exp1} 
       {\hat v_{\rm r}} = \pm \sqrt{- {\rm W}_{k} (x)}
     \end{equation}
   is the general solution of the equation of motion that passes through the point $({\hat r}',\theta{},{\hat v_{\rm r}}')$,
   with the argument function of the W function
     \begin{equation} \label{arg-func-x} 
       x\,({\hat r},\theta) = - {\rm e}^{- f\,({\hat r},\theta; {\hat r}',{\hat v_{\rm r}}')}\, .
     \end{equation}
   Since the argument of the W function in Eq.~(\ref{vr-gensol-exp1}) is always real and negative,
   it is guaranteed that the argument of the square root
   never becomes negative, and hence the solution is always real.  

   Inserting $f\,({\hat r},\theta; {\hat r}',{\hat v_{\rm r}}')$ from Eq.~(\ref{func-f}) into Eq.~(\ref{arg-func-x}) yields
     \begin{eqnarray} \label{x-gensol}
     x\,({\hat r},\theta; {\hat r}',{\hat v_{\rm r}}')\!\! & = & \!\! - \left( \frac{\hat r'}{\hat r} \right)^4
                    {\hat v_{\rm r}}'^{2} \exp \left[ - {\hat v}_{\rm rot}^{2} \, \left( \frac{1}{{\hat r}'^2} - \frac{1}{{\hat r}^2}\right)
		     -2\, {\hat v}_{\rm crit}^{2}\, \left( \frac{1}{\hat r} - \frac{1}{\hat r'} \right) \right.
		       \nonumber\\
                  &  &  \hspace{-4ex} \left. -\frac{2}{\hat r_{0}} \frac{\hat g_{0}}{\delta\left(1+\gamma\right)} 
                     \left(  \left( 1 - \frac{\hat r_{0}}{{\hat r}^{\delta}}\right)^{1+\gamma} - 
		        \left( 1 - \frac{\hat r_{0}}{{\hat r}'^{\delta}}\right)^{1+\gamma}\right) - {\hat v_{\rm r}}'^{2} \right] 
    \end{eqnarray}
    the general expression for the argument function $x$ depending on the parameters $({\hat r}',\theta{},{\hat v_{\rm r}}')$.
    
    Thus, for the trans-sonic case of a stellar wind or general accretion flow, 
    where ${\hat r}'={\hat r}_{\rm c}\,(\theta)\equiv{}{\hat r}_{\rm s}\,(\theta)$
    and ${\hat v_{\rm r}}'={\hat v_{\rm c}}\equiv{}1$,
    the analytical solution is 
     \begin{equation} \label{vr-windsol} 
       {\hat v_{\rm r}}\,({\hat r},\theta) = \pm \sqrt{- {\rm W}_{k} (x\,({\hat r},\theta))} 
     \end{equation}
    with
      \begin{eqnarray} \label{x-windsol} 
     x\,({\hat r},\theta) & = &  - \left( \frac{\hat r_{\rm c}}{\hat r} \right)^4 \, 
                     \exp \left[ - {\hat v}_{\rm rot}^{2} \, \left( \frac{1}{{\hat r_{\rm c}}^2} - \frac{1}{{\hat r}^2}\right)
		     -2\, {\hat v}_{\rm crit}^{2}\, \left( \frac{1}{\hat r} - \frac{1}{\hat r_{\rm c}} \right) \right.
		       \nonumber\\
		   &  & \left. -\frac{2}{\hat r_{0}} \frac{\hat g_{0}}{\delta\left(1+\gamma\right)} \,
                       \left(  \left( 1 - \frac{\hat r_{0}}{{\hat r}^{\delta}}\right)^{1+\gamma} 
		        - \left( 1 - \frac{\hat r_{0}}{{\hat r_{\rm c}}^{\delta}}\right)^{1+\gamma} \right) - 1 \right] \, .
     \end{eqnarray}
     
      We are only interested in the possible two real values of W($x$), the \mbox{$k=0, -1$}--branches
      in Eq.~(\ref{vr-gensol-exp1}) or Eq.~(\ref{vr-windsol}), where $x$ is real and \mbox{$-1/e\le{}x<0$}.
      The branch point at $x=-1/e$, where these two branches meet, 
      corresponds to the critical point ${\hat r}_{\rm c}$, 
      where the velocity in Eq.~(\ref{vr-windsol}) becomes \mbox{${\hat v_{\rm r}}={\hat v_{\rm r}}({\hat r_{\rm c}})$}$\equiv{}1$.      
      Depending on which branch of W one is approaching this point $x=-1/e$, one obtains a different shape of the 
      ${\hat v_{\rm r}}\,({\hat r},\theta)$--curve, i.e. a stellar wind or a collapsing system.
  
      However, to determine which of the two branches to choose 
      at a certain range of radius between \mbox{$[1, {\hat r}_{\rm c}]$} and \mbox{$[{\hat r}_{\rm c}, \infty )$}
      as to guarantee a continuous, monotonically increasing,
      and smooth trans-sonic flow (as, e.g., in the case of a stellar wind),
      one needs to examine the behaviour of the argument function $x({\hat r},\theta)$ of $W$, in Eq.~(\ref{x-windsol}),
      with radius ${\hat r}$ at a given co-latitude $\theta$.
      Then, the argument func\-tion \mbox{$x({\hat r},\theta)$} 
      decreases monotonically from the stellar radius ${\hat r}=1$ (with value of nearly zero) to its minimum at 
      ${\hat r}={\hat r}_{\rm c}$ with $x=-1/e$ to afterwards increase monotonically again.
 
    Finally, a detailed investigation (cf.~Paper I) yields
    the following 
    amount of the radial velocity component (i.e.~the axisymmetric two-dimensional
    trans-sonic analytical solution of our equation of motion for a rotating and expanding or collapsing 
    system) for a given latitude, i.e. polar angle $\theta$
    \begin{itemize}
    \item[(a)] for the case of a stellar wind,
        and Eq.~(\ref{vr-windsol}) (and the positive sign in front of the root) 
	with the argument function in Eq.~(\ref{x-windsol}),
	choosing the branch
	\begin{equation} \label{k-branch-wind}
 	 k =
          \left\{  \begin{array}{ccc}
           \,\,\, 0 & \mbox{for} & 1 \leq {\hat r} \leq {\hat r_{\rm c}}\,(\theta) \\
           -1 & \mbox{for} & {\hat r} > {\hat r_{\rm c}}\,(\theta) 
                       \end{array}
          \right. 
	\end{equation}
	of the W function at a certain radius ${\hat r}$,
	and 
    \item[(b)] in case of a general accretion flux, as well,
        by Eq.~(\ref{vr-windsol}) (but now with the negative sign in front of the root) 
	and the argument function in Eq.~(\ref{x-windsol}),
	choosing the branch	   
  	\begin{equation} \label{k-branch-accretion}
 	 k =
          \left\{  \begin{array}{ccc}
           -1 & \mbox{for} & 1 \leq {\hat r} \leq {\hat r_{\rm c}}\,(\theta) \\
          \,\,\, 0 & \mbox{for} & {\hat r} > {\hat r_{\rm c}}\,(\theta) 
                       \end{array}
          \right. 
	\end{equation}   
         depending on the location ${\hat r}$, where
    \item[(c)] in the special cases of a thermal wind or collapsing system like a
        collapsing protostellar cloud,
	the argument function simplifies to
	\begin{equation}
           x\,({\hat r},\theta)\! =\! -\left( \frac{\hat r_{\rm c}}{\hat r} \right)^4
                     \exp \left[ - {\hat v}_{\rm rot}^{2} \left( \frac{1}{{\hat r_{\rm c}}^2} - \frac{1}{{\hat r}^2}\right)
		     -2\, {\hat v}_{\rm crit}^{2} \left( \frac{1}{\hat r} - \frac{1}{\hat r_{\rm c}} \right) -1 \right]
	\end{equation}
	with ${\hat r_{\rm c}}$ given by Eq.~(\ref{rc-thermalwind}),   
	while choosing the same branches of W as mentioned above in case (a), or (b) respectively,
	and the appropriate sign in Eq.~(\ref{vr-windsol}) (in front of the root).    
     \end{itemize}
     Accordingly, one obtains different expressions for the density distribution, by Eq.~(\ref{rho}) 
     that depends on ${\hat v}_{\rm r}$, which is therefore also piece-wise defined for these different ranges of radius ${\hat r}$.  
     
     In addition to these two critical solutions (type 1 and 2, cf.~numbering in Fig.~\ref{sol-topology}),
     already discussed, 
     that pass through the critical point (i.e. sonic point), all the other four types of solutions were obtained from our general
     velocity law, Eq.~(\ref{vr-gensol-exp1}) with Eq.~(\ref{x-gensol}),
     choosing the following branches of the W function and 
     values of \mbox{(${\hat r}',\theta,{\hat v_{\rm r}}'$)}, for the point we demand the solution to go through:
     \begin{itemize}
     \item Type-3: Subsonic (subcritical) solutions
                    \begin{displaymath}
                     k=0\, , \quad  {\hat r}'= {\hat r}_{\rm c}\,(\theta) , \quad {\hat v_{\rm r}}'<1
                    \end{displaymath}
     \item Type-4: Supersonic (supercritical) solutions
                    \begin{displaymath}
                     k=-1\, , \quad  {\hat r}'= {\hat r}_{\rm c}\,(\theta) \, , \quad {\hat v_{\rm r}}'>1
                    \end{displaymath}
     \item Type-5 and 6: Double-valued solutions 
                     \begin{displaymath}
                     k=0 \quad \mbox{and} \quad k=-1 \quad \mbox{for}
                    \end{displaymath}
                     \begin{displaymath}
                     {\hat r}'\, , {\hat v_{\rm r}}' = \mbox{arbitrarily,\, where} \quad {\hat r}'\neq{}{\hat r}_{\rm c}\,(\theta) \, , 
		     \quad {\hat v_{\rm r}}'\neq{}1\, .
                    \end{displaymath}
     \end{itemize}
      Type-3 solutions are everywhere subsonic
      (choosing only the principal branch, $k=0$, of the W function).
      Those of type 4 are everywhere supersonic
      (selecting only the $k=-1$-branch in the velocity law),
      and those of type 5 and 6 are double-valued,
      composed of both the $k=0$ and $k=-1$-branch, below and above the sonic line, respectively.
       In this connection, the two sub- and supersonic pairs of curves of this last mentioned types,
      subdivided into \mbox{(5a, 6a)} and \mbox{(5b, 6b)} in Fig.~\ref{sol-topology}, belong together.
      They are fixed, not only by the same chosen branch of W in Eq.~(\ref{vr-gensol-exp1}),
      but also by the same selected parameters for the solution through the identical given point 
      \mbox{(${\hat r}',\theta{},{\hat v_{\rm r}}'$)}.
 
     Subsequently, we derive an analytical expression
     for the wind solution in the supersonic approximation (that 
     is only valid in the supersonic region and is not supposed to be applied to the 
     subsonic region,
     where it even becomes imaginary, particularly in our wind model in the range of $0<{\hat r}\lesssim{}{\hat r}_{\rm s}$).
     The reasons for the necessity of deriving this approximated solution are given in our previous paper
     (Paper I) for the solution of a spherical wind.

   \subsubsection{Approximated solution of the equation of motion}  \label{sect-approxsol-eq-mot}

    By neglecting the pressure term $1/\rho\,(\partial{}\rho/{}\partial{\hat r})$ in the equation of motion~(\ref{Eq-of-mot1-nd}), 
    which is a good approximation
     for the stellar wind solution in the supersonic region with ${\hat r}>{\hat r}_{\rm c}\,(\theta)\equiv{}{\hat r}_{\rm s}\,(\theta)$,
     Eq.~(\ref{Eq-of-mot3}) becomes
    \begin{equation}  \label{Eq-of-mot-approx}  
          {\hat v_{\rm r}}\, \frac{\partial}{\partial\,{\hat r}}\, {\hat v_{\rm r}} =
               \frac{{\hat v_{\rm rot}}^{2}}{{\hat r}^3}  -\frac{{\hat v_{\rm crit}}^{2}}{{\hat r}^{2}}
                + \frac{\hat g_{0}}{\hat r^{1+\delta\,\left( 1+\gamma\right)}}\, 
		\left({\hat r}^{\delta} - {\hat r_{0}}\right)^{\gamma}
    \end{equation}
   at the given latitude of interest.
   This simplified equation of motion can again be solved by first integrating the left hand side over
    ${\hat v_{\rm r}}$, and then integrating the right hand side over ${\hat r}$, separately,
    which yields   
         \begin{equation} \label{derive-v-sol-approx} 
     {\hat v_{\rm r}}^{2} =  - \frac{{\hat v_{\rm rot}}^{2}}{{\hat r}^2} + 2\frac{{\hat v}_{\rm crit}^{2}}{\hat r}
                  + \frac{2}{\hat r_{0}}\,\frac{\hat g_{0}}{\delta\,\left(1+\gamma\right)}
		   \left( 1 - \frac{\hat r_{0}}{{\hat r}^{\delta}}\right)^{1 + \gamma} + C \, .
         \end{equation} 
   To determine the integration constant $C$, we assume a boundary condition
   \mbox{${\hat v_{\rm r}}\,({\hat r'}) \approx 0$}
   for the wind velocity at radius ${\hat r'}\,(\theta)={\hat r_{0}}^{1 / \delta}$ and polar angle $\theta$,
   close to the stellar photosphere, i.e.
    \begin{equation}
    C\,({\hat r}'\!=\!{}{\hat r_{0}^{1 / \delta},{\hat v_{\rm r}}'\!=\!{}0) = -2\, \frac{{\hat v}_{\rm crit}^{2}}{{\hat r_{0}}^{1 / \delta}}}
                                                                               + \frac{{\hat v_{\rm rot}}^{2}}{{\hat r_{0}}^{2 / \delta}} \, .
    \end{equation}
    Thus, from Eq.~(\ref{derive-v-sol-approx}), the approximated wind solution reads
    \begin{eqnarray} \label{vwind-sol-approx} 
     {\hat v_{\rm r}}\,({\hat r},\theta) & = & \left[ \frac{2}{\hat r_{0}}\,\left( {\hat v}_{\rm crit}^{2} 
                      \left( \frac{\hat r_{0}}{\hat r} - {\hat r_{0}}^{1 - 1 / \delta} \right) 
                    + \frac{\hat g_{0}}{\delta \left( 1+\gamma\right)} \left( 1 - \frac{\hat r_{0}}{{\hat r}^{\delta}}\right)^{1 + \gamma}\right)
		      \right.
		    \nonumber\\  
		& & \left. + {\hat v_{\rm rot}}^{2}\, \left( \frac{1}{{\hat r_{0}}^{2 / \delta}} - \frac{1}{{\hat r}^2} \right)
                      \right]^{1/2} \, ,
    \end{eqnarray}
    which can be expressed without the W function.

    \subsubsection{Comparison with the $\beta$ velocity law}  \label{sect-comp-beta-law}

   Eq.~(\ref{vwind-sol-approx}) yields a terminal velocity ${\hat v}_{\infty}$ (as ${\hat r}\rightarrow{}\infty$) of
     \begin{equation} \label{vinf-law} 
      {\hat v}_{\infty}\,(\theta) = \sqrt{ \frac{2}{\hat r_{0}}\,\left( \frac{{\hat g_{0}}}{\delta \left(1 + \gamma\right)}
                                   -  {\hat v}_{\rm crit}^{2}\, {\hat r_{0}}^{1 - 1 / \delta}  \right)
				   +  \frac{{\hat v_{\rm rot}}^{2}}{{\hat r_{0}}^{2 / \delta}} } \, ,
     \end{equation}
    dependent on angle $\theta$,
    which is now comparable to the ${\hat v}_{\infty}$ parameter in the (so-called) 
    $\beta$ velocity law (cf. Castor \& Lamers \cite{castor79}; CAK)
    \begin{equation} \label{beta-law} 
     {\hat v_{\rm r}}\,({\hat r}) = {\hat v}_{\infty} \,  \left( 1 - \frac{\hat r_{0}'}{\hat r}\right)^{\beta} 
    \end{equation}
    for a given latitude of interest.
    To be able to compare the $\gamma$ (and $\delta$) parameter in our wind law with the exponent in the
    $\beta$ law (as we use $\beta$ as an input parameter in our model atmosphere calculations),  
   we express our line acceleration parameter ${\hat g_{0}}$ in terms of ${\hat v}_{\infty}$
   by means of Eq.~(\ref{vinf-law}), i.e.
   \begin{equation} \label{g0-vinf-rel} 
     {\hat g_{0}} = \left( \left( \frac{{\hat v}_{\infty}^2}{2} - \frac{{\hat v_{\rm rot}}^{2}}{{\hat r_{0}}^{2 / \delta{} + 1}} \right)
                    + \frac{{\hat v}_{\rm crit}^{2}}{{\hat r_{0}}^{1/\delta}}\right)\,
                    {\hat r_{0}}\, \delta \left( 1 + \gamma \right)\, ,
   \end{equation}
   and insert it into Eq.~(\ref{vwind-sol-approx}), as to obtain
   \begin{eqnarray} \label{vwind-sol-approx2} 
    {\hat v_{\rm r}}\,({\hat r},\theta) & = & \left[ \frac{2}{\hat r_{0}}\,\left( {\hat v}_{\rm crit}^{2} 
                      \left( \frac{\hat r_{0}}{\hat r} - {\hat r_{0}}^{1 - 1 / \delta} \right) 
		      \right. \right.
		      \nonumber\\   
            & & \left. + \left( \left( \frac{\hat r_{0}}{2} {\hat v}_{\infty}^2 - \frac{{\hat v_{\rm rot}}^{2}}{{\hat r_{0}}^{2 / \delta}} \right)
	 	    + {\hat v}_{\rm crit}^{2}\, {\hat r_{0}}^{1 - 1 / \delta} \right)
		                \left( 1 - \frac{\hat r_{0}}{{\hat r}^{\delta}}\right)^{1 + \gamma} \right)
	       	     \nonumber\\  
            & &  \left. + {\hat v_{\rm rot}}^{2}\, \left( \frac{1}{{\hat r_{0}}^{2 / \delta}} - \frac{1}{{\hat r}^2} \right)
                     \right]^{1/2} \, ,
    \end{eqnarray}
   which now depends on ${\hat v}_{\infty}\,(\theta)$, ${\hat v}_{\rm rot}\,(\theta)$, $\gamma\,(\theta)$, $\delta\,(\theta)$ 
   and ${\hat r_{0}}\,(\theta)$.
   
   Since $\delta$ is of the order of 1, as is ${\hat r_{0}}$,
   one can approximate the expression ${\hat r_{0}}^{1 - 1/ \delta}$, in Eq.~(\ref{vwind-sol-approx2}),
   as 1. 
   Furthermore, as our line parameter ${\hat r_{0}}$  
   is defined as the parameter for which the line acceleration becomes zero at radius ${\hat r}={\hat r_{0}}^{1 / \delta}$
   (cf.~Eq.~\ref{line-acc-term1}), this radius is very close to the radius ${\hat r_{0}}'$
   in the $\beta$--law (in Eq.~\ref{beta-law}),
   where the wind velocity is assumed to be zero, i.e. ${\hat r_{0}}'\approx{}{\hat r_{0}}$.
   
   Then, we can set the velocity in Eq.~(\ref{beta-law}) equal to our velocity law in Eq.~(\ref{vwind-sol-approx2}),
   to search for a relationship between the parameters $\beta$, $\gamma$ and $\delta$,
   which yields (for ${\hat v}_{\rm rot}\,(\theta{})\ll{\hat v}_{\infty}\,(\theta{})$)
   \begin{equation}
    \left( 1 - \frac{\hat r_{0}}{\hat r}\right)^{2\beta - 1} \stackrel{!}{\approx}  - b + \left( 1 + b \right)\, 
    \frac{\displaystyle \left( 1 - \frac{\hat r_{0}}{{\hat r}^{\delta}}\right)^{1+\gamma}}
                                     {\displaystyle \!\!\!\!\! \left( 1 - \frac{\hat r_{0}}{\hat r}\right)}
   \end{equation}
   with
   \begin{displaymath}
    b := \frac{2}{\hat r_{0}} \left(\frac{{\hat v}_{\rm crit}}{{\hat v}_{\infty}}\right)^2 \, ,
   \end{displaymath}
   analogous to the one-dimensional case of a spherical wind (cf. Paper I).
   
   Herein, for large radii ${\hat r}$ (and especially for small values of $b$, e.g.~$b\lesssim{}0.1$ for an O-V-star),
   the right hand side can be approximated by the last fraction only,
   which leads to
   \begin{equation} \label{beta-gam-del-rel1}
   \left( 1 - \frac{\hat r_{0}}{\hat r}\right)^{2\beta} \stackrel{!}{\approx} \left( 1 - \frac{\hat r_{0}}{{\hat r}^{\delta}}\right)^{1+\gamma}\, ,
   \end{equation}
   or equivalently
   \begin{equation} \label{beta-gam-del-rel}
   \frac{2 \beta}{1+\gamma} \approx \frac{\displaystyle \log \left( 1 - \frac{\hat r_{0}}{{\hat r}^{\delta}}\right)}
                                                             {\displaystyle \log \left( 1 - \frac{\hat r_{0}}{\hat r}\right)} \, . 
   \end{equation}
   Next, the right hand side in Eq.~(\ref{beta-gam-del-rel}) can be approximately set to 1,
   for values of $\delta\longrightarrow{}1$ or generally for smaller distances from the central star
   in the supersonic region as ${\hat r}\longrightarrow{}{\hat r_{0}}^{1 / \delta}\approx{}1$.
   This results the same relationship between $\beta$ and $\gamma$ as for a non-rotating spherical wind (see Paper I)
   \begin{equation} \label{beta-gam-rel}
    2\, \beta{}\,(\theta{})  \approx  1 + \gamma{}\,(\theta{})   \, ,
   \end{equation}
   independent of $\delta$, that is
   valid for the previously mentioned values for $\delta$ at smaller radii ${\hat r}$.
   It also applies at very large distances ${\hat r}\gg{}1$, since then, the numerical values inside the brackets of Eq.~(\ref{beta-gam-del-rel1}) 
   are close to 1 and this equation is fulfilled for any value of the exponents $\beta$ and $\gamma$ in all cases.
   Only for intermediate distances from the star at lower values of $\delta$ (not close to 1),
   the relationship between $\beta$ and $\gamma$ is possibly not well approximated by
   Eq.~(\ref{beta-gam-rel}).

   \subsubsection{Fitting formula for the line acceleration}   \label{sect-fit-form-gline}
   
   Thus, we can provide another expression for the line acceleration (in Eq.~(\ref{line-acc-term1})),
   now dependent on ${\hat v}_{\rm rot}\,(\theta{})$ and on the parameters 
   ${\hat v}_{\infty}$, $\gamma$ (or $\beta$ equivalently), $\delta$ and ${\hat r_{0}}$
   (by eliminating ${\hat g_{0}}$ using Eq.~(\ref{g0-vinf-rel})):
    \begin{equation} \label{line-acc-term3a}
    {\hat g_{\rm rad}^{\rm line}} ({\hat r},\theta{}) 
    = \left( \left( \frac{{\hat v}_{\infty}^2}{2} - \frac{{\hat v_{\rm rot}}^{2}}{{\hat r_{0}}^{2 / \delta{} + 1}} \right)
                    + \frac{{\hat v}_{\rm crit}^{2}}{{\hat r_{0}}^{1/\delta}}\right)
                    {\hat r_{0}} \, \frac{\delta \left( 1 + \gamma \right)}{\hat r^{1+\delta\,\left( 1+\gamma \right)}}\, 
	            \left({{\hat r}^{\delta}} - {\hat r_{0}}\right)^{\gamma} \, .
    \end{equation}
    This non-linear expression can then be used as fitting formula and applied to
    the results from a numerical calculation of ${\hat g_{\rm rad}^{\rm line}}\,({\hat r}_{i},\theta{})$
    for discrete radial grid points ${\hat r}_{i}$ at a given latitude,
    in order to determine the line acceleration parameters $\gamma\,(\theta{})$ (or equivalently $\beta\,(\theta{})$
    by means of Eq.~(\ref{beta-gam-rel})),
    $\delta\,(\theta{})$, ${\hat r_{0}}\,(\theta{})$
    and the terminal velocity ${\hat v_{\infty}}\,(\theta{})$, cf. Fig.~\ref{pic-gline-fit}.

   \subsubsection{Physical interpretation of the equation of critical radius}  \label{sect-interpret-eq-of-critrad}

   Through the use of the exact wind solution, by using Eq.~(\ref{crit-rad-eq}), valid for the critical radius ${\hat r}_{\rm c}\,(\theta)$
   at latitude with polar angle $\theta$,
   we can solve for the line acceleration parameter ${\hat g}_{0}$
   and insert it into Eq.~(\ref{vinf-law}) from the approximated wind solution,
   to provide another expression for the terminal velocity at $\theta$, 
   depending on the location of the critical (i.e. sonic) point
	  \begin{eqnarray} \label{v-inf-rc}
	{\hat v}_{\infty}\,(\theta) & = & \left[  \frac{2}{\hat r_{0}} \left\lbrace  
         \left( \frac{ {\hat r}_{\rm c}^{\delta} }{ {\hat r}_{\rm c}^{\delta} - {\hat r_{0}}}\right)^{\gamma} 
	   \frac{ {\hat r}_{\rm c}^{\delta - 2} }{\delta \left( 1+\gamma\right)} \left( {\hat v}_{\rm crit}^{2}\,{\hat r}_{\rm c} 
	  - 2\, {\hat r}_{\rm c}^{2} - {\hat v_{\rm rot}}^{2} \right)  \right. \right.
	  \nonumber\\ 
	  & & \left.  - {\hat v}_{\rm crit}^{2} {\hat r_{0}}^{1 - 1/ \delta} \bigg\} 
	       + \frac{{\hat v_{\rm rot}}^{2}}{{\hat r_{0}}^{2 / \delta}}  \right]^{1/2} \, .
	  \end{eqnarray}
    Or vice versa,
    by Eq.~(\ref{crit-rad-eq}), the location of the critical point
    (through which the exact analytical wind solution of our Eq. of motion EOM~(\ref{Eq-of-mot3}) passes) 
    is mainly determined, on the one hand, by the given terminal velocity $v_{\infty}$, via the line acceleration parameter
    ${\hat g_{0}}$.
    On the other hand, the position of ${\hat r_{\rm c}}\,(\theta)$ must also be dependent on the given minimum velocity $v_{\rm in}\,(\theta)$
    at the inner boundary radius $R_{\rm in}\,(\theta)$, where the velocity solution passes.
    This inner velocity $v_{\rm in}$ follows indirectly from the other remaining line acceleration or wind parameters
    $\gamma$, $\delta$ (which make up the shape of the velocity curve) and especially ${\hat r_{0}}$
    (where the value of the latter parameter is determined by the radius ${\hat r_{0}}^{1/\delta}$ at which $g_{\rm rad}^{\rm line}$ is zero).
      
    Since the inner boundary condition of the velocity $v_{\rm in}\,(\theta)$ is connected to the mass-loss rate $\dot{M}\,(\theta)$,
    through the equation of mass continuity by Eq.~(\ref{Def-Mdot(th)}) and the given density at the inner boundary,
    the position of the critical radius ${\hat r_{\rm c}}\,(\theta)$ 
    is uniquely specified by the values of $v_{\infty}\,(\theta)$ and $\dot{M}\,(\theta)$.

    \section{Numerical methods}  \label{sect-num-meth}

    \subsection{Computing the radiative acceleration} \label{sect-num-comp-gline}

    As in Paper I,
    we first calculate the thermal, density and ionisation structure of the wind model by means of the
    non-LTE expanding atmosphere (improved Sobolev approximation) code {\sc ISA-Wind} 
    (de Koter et al.~\cite{dekoter:schmutz}, \cite{dekoter:heap}).
    As a next step, we calculate the radiative acceleration
    as a function of distance by means of a Monte Carlo (MC) code {\sc MC-Wind} 
    (de Koter et al. \cite{dekoter:heap}; Vink et al. \cite{vink:dekoter:lamers}),
    accounting for the possibility that the photons
    can be scattered or eliminated (if they are scattered back into the star).
      The radiative transfer in {\sc MC-Wind} involves multiple continuum and line processes using the Sobolev approximation
    (cf. Mazzali \& Lucy \cite{mazzali:lucy}).
     
    The radiative acceleration of the wind
    at a given constant co-latitude $\theta$ 
    is calculated by following 
    the fate of the large number of photons   
    where the atmosphere is divided into a large number of concentric
    thin shells with radius $r$ and thickness $\ud r$,
    and the loss of photon energy, due to all scatterings that occur within each shell, is determined.
    The total line acceleration per shell summed over all line scatterings in that shell equals 
    (Abbott \& Lucy \cite{abbott:lucy})
     \begin{equation} \label{gline-numerical}
      g_{\rm rad}^{\rm line}\,(r,\theta) = - \frac{1}{\dot{M}\,(\theta{})} \frac{\ud L}{\ud r}\,(r,\theta{}) \, ,
     \end{equation}
    here, (in contrast to the one-dimensional case) referred to a constant polar angle $\theta$,
    where $-\ud L\,(r,\theta{})$ is the rate at which the radiation field loses energy 
    by the transfer of momentum of the photons to the ions of the wind per time interval.
    
    The line list that is used for the MC calculations consists of over $10^5$ of the strongest lines of
    the elements from H to Zn from a line list constructed by Kurucz (\cite{kurucz}).
    Lines in the wavelength region between 50 and 10000 \AA\ are included with ionisation stages up to stage VII.
    The number of photon packets distributed over 
    the spectrum in our wind model, followed from the lower boundary of the atmosphere, is 2--3.5$\times{}10^7$.
    The wind is divided into 90 spherical shells with a large number of narrow shells in the subsonic region 
    and wider shells in the supersonic range.

    \subsection{Computing the mass-loss rate for known stellar \newline and wind parameters}  \label{sect-num-comp-mdot}
   
    By neglecting the pressure term
    and using the expression for the line acceleration per shell (Eq.~(\ref{gline-numerical})),
    an integration of the Eq. of motion~(\ref{Eq-of-mot1-nd}) at a given constant latitude, 
    from stellar radius to infinity, yields (cf. Abbott \& Lucy \cite{abbott:lucy})
     \begin{displaymath}
    \frac{1}{2}\, \dot{M}\,(\theta{}) \, \left( v_{\infty}^2\,(\theta{}) + v_{\rm esc}^2\,(\theta{}) \right) = \Delta L\,(\theta{}) \, ,
     \end{displaymath}
   or equivalently,
    \begin{equation} \label{Mdot-numerical}
     \dot{M}\,(\theta{}) = \frac{2\, \Delta\, L\,(\theta{})}{v_{\infty}^2\,(\theta{}) + v_{\rm esc}^2\,(\theta{})}
     \end{equation}
   (as in Paper I, but here depending on $\theta$),
   where $\Delta{}L\,(\theta{})$ is the total amount of radiative energy extracted per second, summed over all the shells
   (at co-latitude $\theta$).
   This equation is now fundamental for determining mass-loss rates numerically
   from the total removed radiative luminosity, for the prespecified stellar and wind 
   parameters $v_{\rm esc}\,(\theta{})$ and $v_{\infty}\,(\theta{})$, respectively.

   \subsection{The iteration method: determination of $\dot{M}(\theta)$ and the wind parameters} \label{sect-it-method}

    To compute the mass-loss rate and the wind model parameters from a given rotating central star
    with the fixed stellar parameters\footnote{In the general case of a non-spherical rotating star, the stellar parameters
    $L$, $T_{\rm eff}$, $R$, and $\Gamma$ depend on $\theta$, see Sects.~\ref{sect-azimuth-comp-oblateness} and~\ref{sec-grav-dark}.} 
    $L\,(\theta)$, $T_{\rm eff}\,(\theta)$, $R\,(\theta)$, $M$, $\Gamma\,(\theta)$, $V_{\rm rot}$,
    the analogous iterative procedure as in the one-dimensional non-rotating case can be applied (cf.~Paper I).
    However, the whole iteration cycle here has to be performed separately for each given co-latitude $\theta$ of interest:
    \begin{enumerate}
     \item By keeping the stellar and wind parameters $\dot{M}_{n}(\theta{})$, $v_{{\infty}_{n}}(\theta{})$, $\beta_{n}(\theta{})$ 
           variable throughout our iteration process, 
	   we use arbitrary (but reasonable) starting values $\dot{M}_{-1}$, $v_{{\infty}_{-1}}$, $\beta_{-1}$  
           in iteration step $n=-1$ (cf. Tables~\ref{tablA1-wind-non-dist} -- \ref{tablA3-wind-500kms}).
     \item For these input parameters, a model atmosphere is calculated with {\sc ISA-Wind} for constant co-latitude $\theta$.
           The code yields the thermal structure, the ionisation and excitation structure, and the population of
	   energy levels of all relevant ions.
           Then, the radiative acceleration $g_{\rm rad}^{\rm line}\,(r_{i},\theta{})$ is calculated
           for discrete radial grid points $r_{i}$ at polar angle $\theta$
	   with {\sc MC-Wind} and Eq.~(\ref{gline-numerical}).
	   In addition, an improved estimate for the mass-loss rate $\dot{M}_{n}^{\rm out}(\theta{})$ is obtained by Eq.~(\ref{Mdot-numerical}),
	   which can be used as a new input value for the next iteration step.
	   Moreover, one obtains a new output value for the sonic radius ${\hat r}_{{\rm s}_{n}}(\theta{})$
           (which has to be equal to the critical radius ${\hat r}_{\rm c}(\theta{})$ of our wind theory).
     \item To determine the improved line acceleration parameters $\gamma_{n}(\theta{})$ (or equivalently $\beta_{n}(\theta{})$),
           ${\delta}_{n}(\theta{})$ and ${\hat r_{{0}_{n}}}(\theta{})$ for the considered latitude,
	   Eq.~(\ref{line-acc-term3a}) (together with Eq.~(\ref{beta-gam-rel})) is used as the fitting formula to apply 
           to the numerical
	   results for $g_{\rm rad}^{\rm line}\,(r_{i},\theta{})$, cf. Fig.~\ref{pic-gline-fit}.
     \item By applying Eq.~(\ref{v-inf-rc}) and inserting the current values of parameters $\gamma_{n}$, 	   
	    ${\delta}_{n}$ and ${\hat r_{{0}_{n}}}$, as well as the current sonic radius ${\hat r}_{{\rm s}_{n}}$ for ${\hat r}_{\rm c}$,
	    we obtain a new approximation of the terminal velocity $v_{{\infty}_{n}}(\theta{})$, i.e.
	  \begin{eqnarray} \label{v-inf-next}
      v_{{\infty}_{n}}(\theta) \!\! & = & \!\! a \left[ \frac{2}{\hat r_{{0}_{n}}} \left( 
           \left( \frac{ {\hat r}_{{\rm s}_{n}}^{{\delta}_{n}} }{ {\hat r}_{{\rm s}_{n}}^{{\delta}_{n}} - {\hat r_{{0}_{n}}}}\right)^{\gamma_{n}} 
	     \!\!\! \frac{ {\hat r}_{{\rm s}_{n}}^{{\delta_{n}} - 2} }{\delta_{n} \left( 1+\gamma_{n}\right)} 
	          \left( {\hat v}_{\rm crit}^{2}\, {\hat r}_{{\rm s}_{n}}  
	           - 2\, {\hat r}_{{\rm s}_{n}}^{2} - {\hat v}_{\rm rot}^{2}\right)   \right. \right.
		\nonumber\\   
		 & &  \left.  - {\hat v}_{\rm crit}^{2} {\hat r_{{0}_{n}}}^{1 - 1/ {\delta_{n}}} \Bigg) 
		     + \frac{{\hat v_{\rm rot}}^{2}}{ {\hat r_{{0}_{n}}}^{2 / \delta_{n} }} \right]^{1/2}  \, .
          \end{eqnarray}
     \item With these improved estimates of $\dot{M}_{n}(\theta{})$, $v_{{\infty}_{n}}(\theta{})$, $\beta_{n}(\theta{})$ as new input parameters, 
	   the whole iteration step, defined by items 2 -- 4, is repeated until convergence (at given latitude)
	   is achieved.	   
    \end{enumerate}

    \subsection{The adjustment of the wind formalism to ISA-Wind}  
    
    The {\sc ISA-Wind} code has already been described in detail by de Koter et al. (\cite{dekoter:schmutz}, \cite{dekoter:lamers}).
    Those model assumptions within {\sc ISA-Wind} which affect our wind formalism,
    by using the code in our iteration process, have also been described in Paper I.
    
    To be able to apply the analytical wind solution of our EOM~(\ref{Eq-of-mot3}) to model a stellar wind
    from a given central star (with fixed stellar parameters)
    by using {\sc ISA-Wind} to find numerically the unique solution,
    we had to adjust our wind formalism, i.e. our more accurate EOM, to the assumed EOM and wind velocity structure in {\sc ISA-Wind}.
    The EOM~(\ref{Eq-of-mot3}) (and therefore our exact analytical wind solution) considers (allows) a line acceleration
    throughout the whole wind regime, starting above the radius ${\hat r}_{0}^{1/\delta}$,
    whereas the different EOM in {\sc ISA-Wind} is only solved in the subsonic region by neglecting the line force.
    However then, {\sc ISA-Wind} 'switches on' the line force somewhere below a connection radius ${\hat r}_{\rm con}$  
    in the subsonic region by assuming a $\beta$ velocity law above ${\hat r}_{\rm con}$ in the supersonic region.
    
    This inconsistency through the use of {\sc ISA-Wind} (compared to our model assumptions and solutions) 
    has been eliminated by introducing the parameter ${\hat r}_{0}\,(\theta)$ into our formalism of Sect.~\ref{sect-analytical-theory}
    for which the line radiation force is zero
    at radius ${\hat r}_{0}^{1/\delta}$ at a given co-latitude $\theta$.
    Then, the final value of ${\hat r}_{0}\,(\theta)$, together with the other remaining line acceleration or wind parameters,
    can be determined by fitting Eq.~(\ref{line-acc-term3a}) and the 
    iteration procedure. 
    
    Moreover, a further inconsistency would occur if we applied {\sc ISA-Wind}, developed for non-rotating spherical winds,
    in our iteration method to compute the wind parameters from a rotating star:
    the EOM in {\sc ISA-Wind} for the subsonic region and the assumed $\beta$ velocity law for the supersonic wind region, therein,
    do not consider the additional centrifugal term \mbox{${\hat v_{\rm rot}}^{2}/{\hat r}^3$} in contrast to our more 
    accurate EOM~(\ref{Eq-of-mot3}),
    which we solve for the radial wind velocity from rotating stars.
    Therefore, to compensate for this discrepancy, one has to neglect the centrifugal term
    in our simplified EOM~(\ref{Eq-of-mot-approx}) for the supersonic wind region,
    which results a simpler approximated wind solution and terminal velocity $v_{\infty}\,(\theta)$ than that in Eq.~(\ref{vwind-sol-approx})
    and Eq.~(\ref{vinf-law}), respectively, where ${\hat v_{\rm rot}}$ can be set to zero.           
    Then, this in turn, leads also to a simpler expression for the fitting formula for the line acceleration
    where the ${\hat v_{\rm rot}}$ term in Eq.~(\ref{line-acc-term3a}) vanishes. 
    However, the derivation of the expression for the terminal velocity $v_{\infty}\,(\theta)$, as a function
    of critical radius ${\hat r}_{\rm c}$ (or sonic radius ${\hat r}_{\rm s}$) in Sect.~\ref{sect-interpret-eq-of-critrad},
    yields then almost the same Eq.~(\ref{v-inf-rc}), where only the last \mbox{${\hat v_{\rm rot}}^{2}/{\hat r_{0}}^{2 / \delta}$}-term
    (from the simplified approximated wind solution) vanishes but not
    the \mbox{${\hat v_{\rm rot}}^{2}$}-term that originates from our equation of critical radius, Eq.~(\ref{crit-rad-eq}).
    The latter takes the stellar rotational speed ${\hat v_{\rm rot}}\,(\theta)$ at co-latitude $\theta$ 
    and its influence on the location of the critical point into consideration.
    
    To avoid this inconsistency in our subsequent wind models for rotating O--stars in Sect.~\ref{sect-applic-ostar},
    we have thus applied the following simplified fitting formula
    \begin{equation} \label{simpl-line-acc}
    {\hat g_{\rm rad}^{\rm line}} ({\hat r},\theta{}) 
    = \left( \frac{{\hat v}_{\infty}^2}{2}
                    + \frac{{\hat v}_{\rm crit}^{2}}{{\hat r_{0}}^{1/\delta}}\right)
                    {\hat r_{0}} \, \frac{\delta \left( 1 + \gamma \right)}{\hat r^{1+\delta\,\left( 1+\gamma \right)}}\, 
	            \left({{\hat r}^{\delta}} - {\hat r_{0}}\right)^{\gamma}
    \end{equation}
    (instead of Eq.~(\ref{line-acc-term3a})) to determine gradually the improved line acceleration parameters
    by our iterative procedure for a given constant co-latitude $\theta$,
    whereas the new estimate of the terminal velocity for the next step has been determined by the simplified iteration formula
    \begin{eqnarray} \label{simpl-v-inf-next}
      v_{{\infty}_{n}}(\theta) \!\! & = & \!\! a \left[ \frac{2}{\hat r_{{0}_{n}}} \left( 
           \left( \frac{ {\hat r}_{{\rm s}_{n}}^{{\delta}_{n}} }{ {\hat r}_{{\rm s}_{n}}^{{\delta}_{n}} - {\hat r_{{0}_{n}}}}\right)^{\gamma_{n}} 
	     \!\!\! \frac{ {\hat r}_{{\rm s}_{n}}^{{\delta_{n}} - 2} }{\delta_{n} \left( 1+\gamma_{n}\right)} 
	          \left( {\hat v}_{\rm crit}^{2}\, {\hat r}_{{\rm s}_{n}}  
	           - 2\, {\hat r}_{{\rm s}_{n}}^{2} - {\hat v}_{\rm rot}^{2}\right)   \right. \right.
		\nonumber\\   
		 & &  \left.  - {\hat v}_{\rm crit}^{2} {\hat r_{{0}_{n}}}^{1 - 1/ {\delta_{n}}} \Bigg) \right]^{1/2} \, ,
    \end{eqnarray}
    instead of the generally more accurate Eq.~(\ref{v-inf-next}).
    
    Further, since {\sc ISA-Wind} begins its computations already below (however close to) the stellar (i.e. photospheric) radius,
    all formulae derived in Sect.~\ref{sect-analytical-theory} have been applied with reference to the inner boundary (core) 
    radius $R_{\rm in}\,(\theta)$ from where the numerical calculations of the wind model start.
    Therefore, the dimensionless variable of distance ${\hat r}$ at polar angle $\theta$ (in Sect.~\ref{sect-applic-ostar})
    refers to $R=R_{\rm in}\,(\theta)$.

    \subsection{The chosen boundary values in the wind models} \label{sect-chosen-bound-values}
     
    In our following numerical wind models  
    the inner boundary radius has been chosen constant throughout the whole iteration process for each chosen latitude and star.
    E.g., for the particular case of the wind from the undistorted rotating O--main-sequence star (cf. Table~\ref{table1-par-nodist-star}),
    it is even constantly $R_{\rm in}=$\mbox{$11.757\,R_{\sun}$} at each co-latitude $\theta$ (e.g.~at 0 and $\pi{}/2$),
    situated at a prescribed fixed Rosseland optical depth of about $\tau_{\rm R}=23$.  
    This corresponds then to a photospheric radius of $R_{\rm phot}=11.828\,R_{\sun}$ (at each latitude), 
    defined as where the thermal optical depth is $\tau_{\rm e}=1/\sqrt{3}$,
    and an inner boundary density of $\rho_{\rm in}=1.398\times{}10^{-8}$~g/cm$^{3}$ at $R_{\rm in}$.
    
    Small changes of this particular chosen fixed value for $\tau_{\rm R}$, or corresponding $\rho_{\rm in}\,(\theta)$,
    at each step of the iteration cycle
    (generally dependent on the given co-latitude $\theta$ and stellar rotation $V_{\rm rot}$
    in the case of a distorted star, cf. Table~\ref{table2-par-dist-star})
    would have no effect on the final wind parameters,
    i.e. in particular on the converged values of $\dot{M}\,(\theta)$ and $v_{\infty}\,(\theta)$,
    as already explained in our 1D Paper~I.

    \section{Application: results for rotating O--stars}  \label{sect-applic-ostar}
    
      In this section we apply our theoretical results from Sect.~\ref{sect-analytical-theory}
      and the iterative procedure described in Sect.~\ref{sect-it-method}
      to first compute the stellar wind parameters for a differentially rotating  40\,$M_{\sun}$ O5--V main-sequence star
      with an equatorial rotation speed of $V_{\rm rot} = 300$~km~s$^{-1}$ ($\Omega = 0.42$) and 500~km~s$^{-1}$ ($\Omega = 0.70$), respectively.
      Secondly, we determine the wind parameters for a 60\,$M_{\sun}$ O--giant star rotating with $V_{\rm rot}=300$~km~s$^{-1}$ ($\Omega = 0.55$).  
      In the first instance, we ignore the effects of stellar distortion and gravity darkening in order 
       to be able to compare our procedures against previous models, such as those from FA, before 
      we include the more realistic aspects involving stellar distortion in Sect.~\ref{s_dist}.
  
      \subsection{Rotating O--stars without distortion}
      
%
    \begin{table}
    \begin{center}
    \caption{Stellar and wind parameters for a differentially rotating O5-V main sequence star
    (with $V_{\rm rot}=300$~km~s$^{-1}$ ($\Omega = 0.42$) and 500~km~s$^{-1}$ ($\Omega = 0.70$), respectively, without distortion) at the pole ($\theta{}=0$)
    and the equator ($\theta{}=\pi/2$).$^a$}
    \label{table1-par-nodist-star}
     \begin{tabular}{lcl}
    \hline\hline\noalign{\smallskip} 
     $M$ & = & 40.0 $M_{\sun}$ \\
     $R_{\rm core}$ & = & 11.757 $R_{\sun}$ \\
     $\tau\,(R_{\rm core})$ & = & 22.9 \\
     $\rho\,(R_{\rm core})$ & = & $1.398 \times{}10^{-8}$ g cm$^{-3}$ \\
     $T_{\rm eff}$ & = & 40000 K \\
     $\log\,\left( L / L_{\sun}\right)$ & = & 5.5 \\
     $\Gamma$ & = & 0.214 \\
     Element abundances & = & Solar metallicity \\
     \end{tabular}
     \vspace{0.8cm}
     \begin{tabular}{llll}
   \hline\hline\noalign{\smallskip}
   $V_{\rm rot}$ [km s$^{-1}$] &  \hspace{-4ex} 0, 300, 500 & 300 & 500 \\
   $\theta$ [rad] & 0 & $\pi{}/2$ & $\pi{}/2$        \\
    \hline\noalign{\smallskip} 
    $\log\dot{M}$ [$M_{\sun}/yr$] &  -6.046   &  -6.026   &  -5.937     \\
    $v_{\infty}$ [km s$^{-1}$]    &   3240    &  3086     &  2720       \\
    $\beta$                       &  0.731    &  0.757    &  0.808      \\
    ${\hat g_{0}}$                &  17392    &  17321    &  14984      \\
    $\gamma$                      &  0.462    &  0.515    &  0.616      \\
    $\delta$                      &  0.6811   &  0.716    &  0.727      \\
    ${\hat r_{0}}$ [$r_{0}/R_{\rm core}$] & 1.0014  & 1.0005 & 0.9993 \\
    ${\hat r_{\rm c}}$ [$r_{\rm c}/R_{\rm core}$] & 1.0098 & 1.0094 & 1.0108 \\
    ${\hat r_{\rm s}}$ [$r_{\rm s}/R_{\rm core}$]\,(ISA) &  1.011  & 1.011 & 1.012 \\
    $\left(\rho_{\rm eq}/{}\rho_{\rm p}\right)_{\infty}$ & --  &  1.099  & 1.531 \\
     \hline 
    \end{tabular}
    \end{center}
    \vspace{-1cm}
     \begin{flushleft} 
     $^a$ The results for the wind from the differentially rotating
     central star with the fixed stellar parameters given in the \emph{upper part} of the table.
     The converged values of the wind parameters $v_{\infty}$, $\beta$, ${\hat g_{0}}$, $\gamma$, $\delta$, 
     ${\hat r}_{0}$, and mass-loss rates $\log\,\dot{M}$, together with the critical radius ${\hat r_{\rm c}}$,
     which is equal (i.e. close) to the sonic radius  ${\hat r_{\rm s}}$ (obtained by {\sc ISA-Wind}),
     are displayed in the \emph{lower part} of the table: 
     the results of the first column (at the pole) correspond to the spherical 1D wind model published in Paper I (see Table~5),
     whereas the second and third column represent the converged parameters of the iteration cycles shown
     in Table~\ref{tablA1-wind-non-dist} at the equator for the different stellar rotation speeds of 
     $V_{\rm rot}=$300 and 500~km~s$^{-1}$, respectively. 
     In addition, the terminal ratios $(\rho_{\rm eq}/{}\rho_{\rm p})_{\infty}$ of the wind densities at the equator compared to
     the pole are also calculated. 
     \end{flushleft}
    \end{table}
%
%

%
    \begin{table}
    \begin{center}
    \caption{Stellar and wind parameters for an O--giant rotating with an equatorial speed of
    $V_{\rm rot}=300$~km~s$^{-1}$ and $\Omega = 0.55$ (without distortion) at the pole ($\theta{}=0$)
    and the equator ($\theta{}=\pi/2$).$^a$}
    \label{table1b-par-nodist-star}
     \begin{tabular}{lcl}
    \hline\hline\noalign{\smallskip} 
     $M$ & = & 60.0 $M_{\sun}$ \\
     $R_{\rm core}$ & = & 20.8787 $R_{\sun}$ \\
     $\tau\,(R_{\rm core})$ & = & 23.2 \\
     $\rho\,(R_{\rm core})$ & = & $6.568 \times{}10^{-9}$ g cm$^{-3}$ \\
     $T_{\rm eff}$ & = & 40000 K \\
     $\log\,\left( L / L_{\sun}\right)$ & = & 6.0 \\
     $\Gamma$ & = & 0.449 \\
     Element abundances & = & Solar metallicity \\
     \end{tabular}
     \vspace{0.8cm}
     \begin{tabular}{lll}
   \hline\hline\noalign{\smallskip}
   $\theta$ [rad] & 0 & $\pi{}/2$  \\
    \hline\noalign{\smallskip} 
    $\log\dot{M}$ [$M_{\sun}/yr$] &  -5.361   &  -5.03 $\pm$ 0.03   \\
    $v_{\infty}$ [km s$^{-1}$]    &   3280    &  2880 $\pm$ 25  \\
    $\beta$                       &  0.852    &  0.73 $\pm$ 0.01  \\
    ${\hat g_{0}}$                &  23482    &  9106 $\pm$ 1028  \\
    $\gamma$                      &  0.705    &  0.46 $\pm$ 0.03  \\
    $\delta$                      &  0.798    &  0.46 $\pm$ 0.03  \\
    ${\hat r_{0}}$ [$r_{0}/R_{\rm core}$] & 1.0024  & 1.0011 $\pm$ 0.0004 \\
    ${\hat r_{\rm c}}$ [$r_{\rm c}/R_{\rm core}$] & 1.0157 & 1.0091 $\pm$ 0.0008 \\
    ${\hat r_{\rm s}}$ [$r_{\rm s}/R_{\rm core}$]\,(ISA) &  1.0180  &  1.0160 $\pm$ 0.0003 \\
    $\left(\rho_{\rm eq}/{}\rho_{\rm p}\right)_{\infty}$ & --  &  2.4 $\pm$ 0.2 \\
     \hline 
    \end{tabular}
    \end{center}
    \vspace{-1.2cm}
     \begin{flushleft} 
     $^a$ The results for the wind from the rotating O--giant
     with the fixed stellar parameters given in the \emph{upper part} of the table.
     The converged values of the wind parameters $v_{\infty}$, $\beta$, ${\hat g_{0}}$, $\gamma$, $\delta$, 
     ${\hat r}_{0}$, and mass-loss rates $\log\,\dot{M}$, together with the critical radius ${\hat r_{\rm c}}$,
     which is equal (i.e. close) to the sonic radius ${\hat r_{\rm s}}$ (obtained by {\sc ISA-Wind}),
     are displayed in the \emph{lower part} of the table: 
     the values of the first column (at the pole) are the results of the iteration cycle shown
     in the upper part of Table~\ref{tabl-A1b-giant-wind},
     whereas the mean values for the equator in the second column were obtained from the parameter values
     of iteration step no 14 and 23, displayed in the lower part of Table~\ref{tabl-A1b-giant-wind}.
     $(\rho_{\rm eq}/{}\rho_{\rm p})_{\infty}$ is the terminal ratio of the wind density at the equator compared to the pole. 
     \end{flushleft}
    \end{table}
\begin{figure*}
\centerline{\hbox{\hspace{0.2cm}\epsfxsize=8.7cm \epsfbox{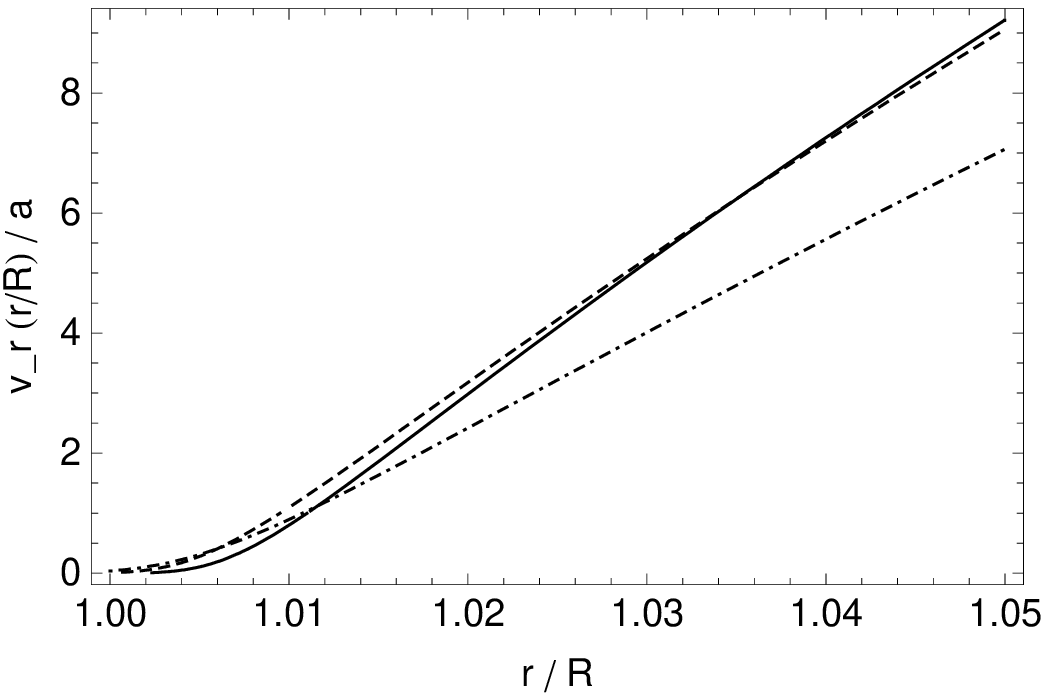}
\hspace{0.2cm} \epsfxsize=9cm \epsfbox{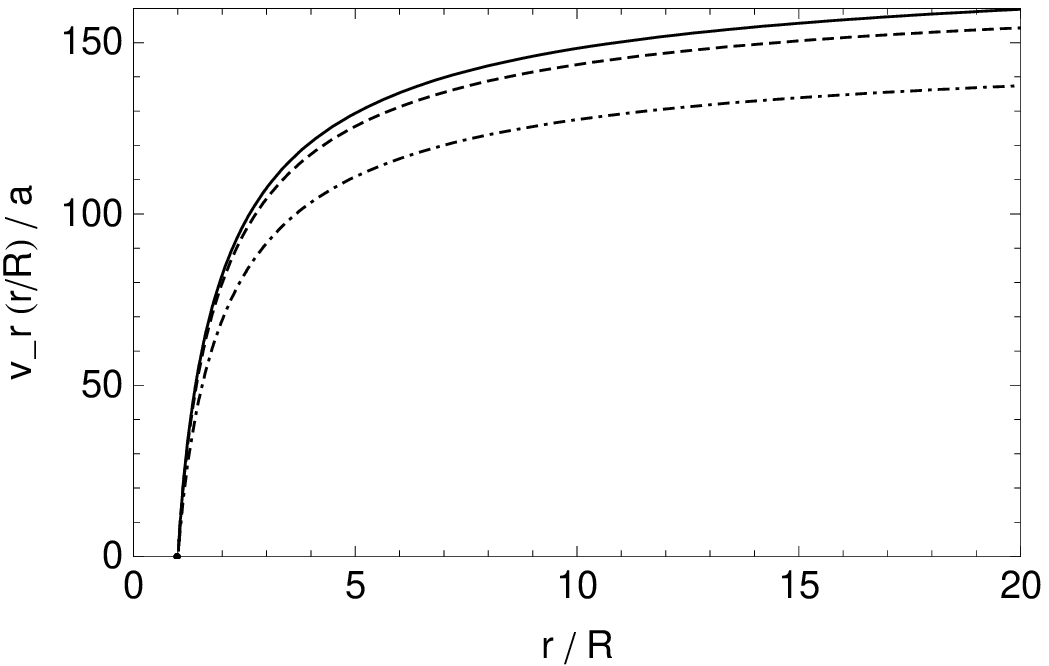}}}
\vspace{0.2cm}
\centerline{\hbox{\hspace{0cm}\epsfxsize=8.7cm \epsfbox{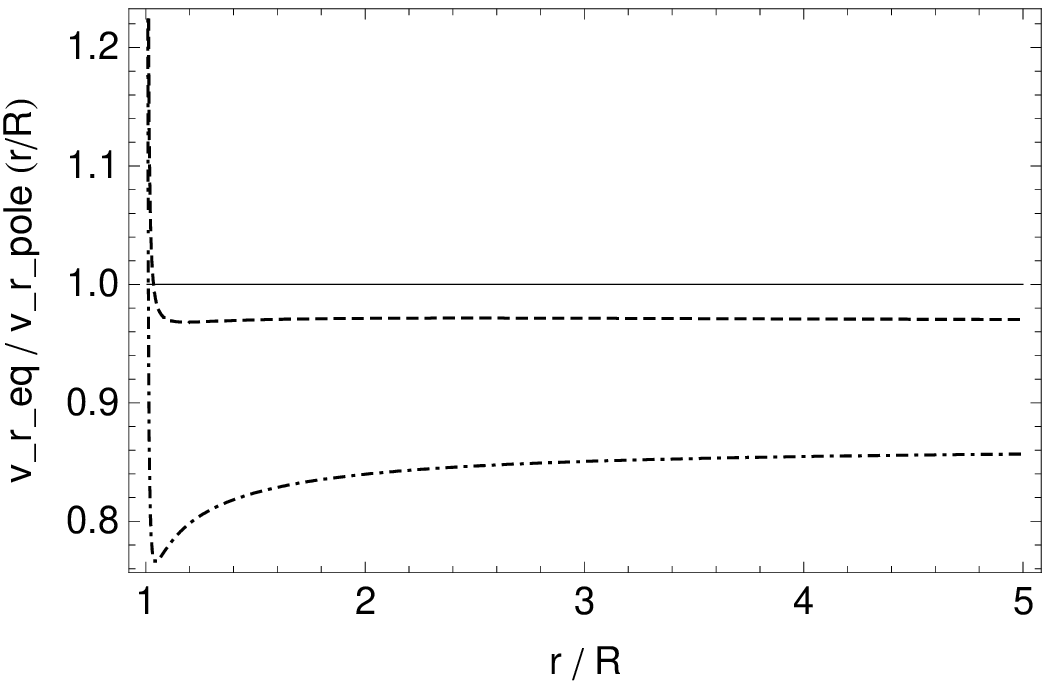}
\hspace{0.2cm} \epsfxsize=8.7cm \epsfbox{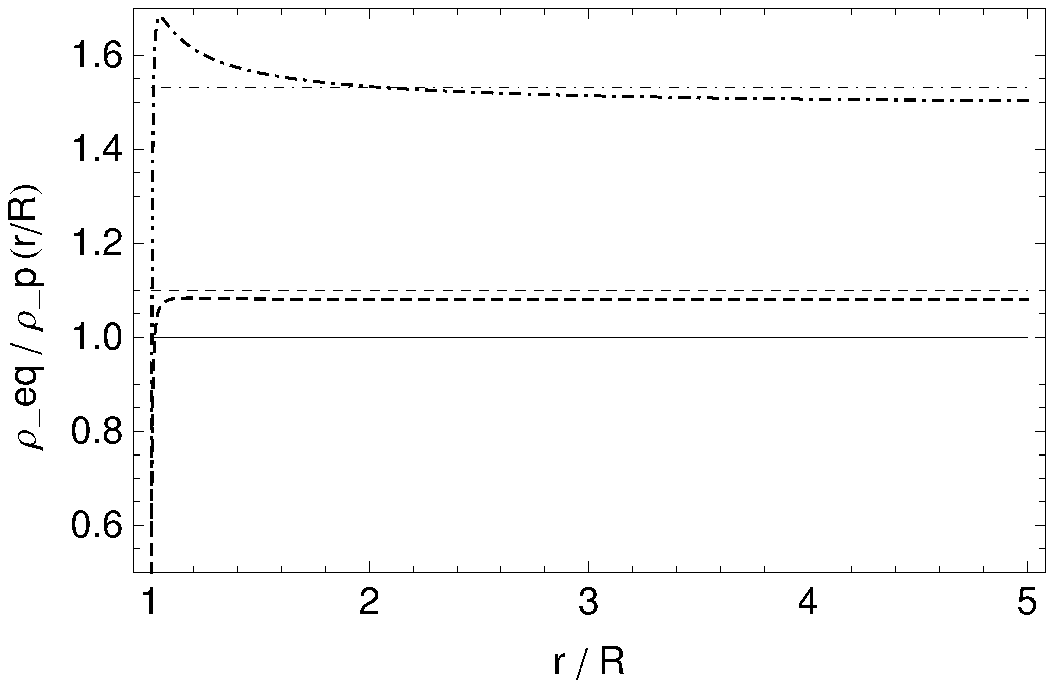}}}
\vfill\parbox[b]{18cm}{\caption[]{
      Model results for the wind from a differentially rotating O5--V main-sequence star \emph{without distortion}
      with an equatorial rotation speed of $V_{\rm rot}=300$~km~s$^{-1}$ (see dashed curves) and 500~km~s$^{-1}$ (see dotted-dashed curves)
      \emph{at the equator} (for $\theta=\pi{}/2$),
      compared to the spherical wind from a corresponding \emph{non-}rotating star ($V_{\rm rot}=0$) which is identical to
      the wind from the rotating star \emph{at the pole} for $\theta=0$ (see solid curves or solid horizontal lines, respectively);
      as for the stellar and wind parameters see Table~\ref{table1-par-nodist-star} in Sect.~\ref{sect-applic-ostar}.
      All diagrams are plotted vs. radial distance ${\hat r}$ in units of the stellar core radius $R=11.757~R_{\sun}$.
      \emph{Upper panel}: the amount of the radial wind velocity component ${\hat v}_{\rm r}\,({\hat r},\theta{})$
      (in units of sound speed $a$) in the subsonic and lower supersonic wind regime (see \emph{left diagram}), and
      in the supersonic region up to stellar distances of ${\hat r}=20$ (see \emph{right diagram}).
      \emph{Lower panel}: the \emph{left diagram} displays the curves of the ratio 
      $v_{\rm r}\,({\hat r},\pi/2)/v_{\rm r}\,({\hat r},0)$ of the equatorial radial wind velocity compared to
      the polar wind velocity and the \emph{right diagram} shows the ratio $\rho\,({\hat r},\pi/2)/{}\rho\,({\hat r},0)$
      of the equatorial wind density in terms of the density at the pole.
      For the pole, these ratios are simply represented by the (solid) constant lines at $1.0$;
      at the equator the density ratios approach (for large stellar distances ${\hat r}$)
      the terminal values represented by the thin horizontal dashed or dotted-dashed line, respectively, in the right diagram.
      } \label{pic-vr-rho-nodist-star} }
\end{figure*}

      The fixed parameters (see e.g. Martins et al. \cite{martins:al} and references therein)
      of the O--main-sequence star are
      given in the upper part of Table~\ref{table1-par-nodist-star},
      whereas the fixed stellar parameters of the 60\,$M_{\sun}$ O--giant are displayed in the upper part of Table~\ref{table1b-par-nodist-star}.
      The iteration steps are provided in the Appendix. The final results are listed in the lower part of 
      Tables~\ref{table1-par-nodist-star} and\,\ref{table1b-par-nodist-star} respectively, and shown graphically 
      for the OV star in Fig.~\ref{pic-vr-rho-nodist-star}. It can be noted that 
      the predicted mass-loss has barely changed from the spherical case for the model rotating with 300 km/s. For the 
      more rapidly rotating 500 km/s model, the equatorial mass-loss rate increases slightly, by $\sim$ 0.1 dex, in rough agreement with
      previous studies, such as FA. Furthermore, the terminal wind velocities also remain relatively constant, although they drop by 
      up to 15\% for the most rapid models. 

      The ratio of 
      equatorial-to-polar density ($\rho_{\rm eq}/\rho_{\rm p}$) reaches a constant value at a sufficient distance from the stellar surface, which 
      follows from Eq.~(\ref{rho})
      \begin{equation}
        \left(\frac{\rho_{\rm eq}}{\rho_{\rm p}}\right)_{\infty} =
        \frac{\dot{M}_{\rm eq}}{\dot{M}_{\rm p}} \,
        \frac{v_{{\infty}_{\rm p}}}{v_{{\infty}_{\rm eq}}}
      \end{equation}
      as illustrated, e.g., in the right diagram of Fig.~\ref{pic-vr-rho-nodist-star} (lower panel).
      We therefore provide this ratio at infinity in all Tables~\ref{table1-par-nodist-star}--\ref{table2-par-dist-star}.

      For the rapid rotator 
      (500 km/s) with the slightly enhanced equatorial mass-loss rate, we find a density contrast ratio of equator to the pole that varies with stellar distance, has 
       a maximum of about 1.7 close to the star, and then approaches a constant value of $\sim$1.5. 
       
      Next we turn our attention to the 60\,$M_{\sun}$ O--giant rotating with $V_{\rm rot}=300$~km~s$^{-1}$. Here we find the effects of rotation 
      to be more pronounced than for the main sequence dwarf. Already at velocities as low as 300 km/s the mass-loss rate 
      at the equator is more than a 
      factor two larger than for the spherical non-rotating case, and the equator-to-pole density contrast ratio is  $(\rho_{\rm eq}/\rho_{\rm p})_{\infty}$ $\simeq$2.4.

    \subsection{Considering oblateness and gravity darkening \newline of the rotating O--main sequence star}
    \label{s_dist}

%
    \begin{table*}
    \begin{center}
    \caption{Stellar and wind parameters for a differentially rotating O5-V main sequence star
    (with $V_{\rm rot}=300$~km~s$^{-1}$ ($\Omega = 0.70$) and 500~km~s$^{-1}$ ($\Omega = 0.92$), respectively) at the pole ($\theta{}=0$)
    and the equator ($\theta{}=\pi/2$),
    considering the stellar distortion and the effects of gravity darkening.$^a$}
    \label{table2-par-dist-star}
     \begin{tabular}{lllll}
   \hline\hline\noalign{\smallskip}
   $V_{\rm rot}$ [km s$^{-1}$] &  \multicolumn{2}{l}{\hspace{6ex}300} & \multicolumn{2}{l}{\hspace{6ex}500} \\
   $\theta$ [rad] & 0 & $\pi{}/2$ & 0 & $\pi{}/2$        \\
    \hline\noalign{\smallskip} 
    $R_{\rm core}\,(\theta)$ [$R_{\sun}$] & 11.757 & 12.870  & 11.757  &  15.477 \\
    $\tau\,(R_{\rm core})$ & 23.0  & 23.5  & 23.8  &  23.4 \\
    $\rho\,(R_{\rm core})$[g cm$^{-3}$] & 1.398 $\times 10^{-8}$ & 1.220 $\times 10^{-8}$ & 1.398 $\times 10^{-8}$ & 8.186 $\times 10^{-9}$ \\
    $T_{\rm eff}\,(\theta)$ [K] & 41242 & 37401 & 43770  &  29695 \\
    $\log\left( L\,(\theta{}) / L_{\sun}\right)$ & 5.56  & 5.47 & 5.66  &  5.23 \\
    $\Gamma\,(\theta)$ & 0.241  &  0.196  & 0.306 & 0.104  \\
    \noalign{\smallskip} \noalign{\smallskip}
    $\log\dot{M}$ [$M_{\sun}/yr$] &  -6.09  &  -6.09 $\pm$ 0.02  &  -6.09 & -6.918 $\pm$ 0.002   \\
    $v_{\infty}$ [km s$^{-1}$]    &   3607  &  3710 $\pm$ 173    &  4755  &  3837 $\pm$ 10       \\
    $\beta$                       &  0.853  &  0.68 $\pm$ 0.02 & 0.999  &  1.048 $\pm$ 0.008     \\
    ${\hat g_{0}}$                &  31563 & 13970 $\pm$ 235  &  64923  &  54979 $\pm$ 609    \\
    $\gamma$                      &  0.707  & 0.35 $\pm$ 0.04 & 0.999  & 1.10 $\pm$ 0.02   \\
    $\delta$                      &  0.898  & 0.43 $\pm$ 0.03 & 0.997  & 0.830 $\pm$ 0.006     \\
    ${\hat r_{0}}$ [$r_{0}/R_{\rm core}\,(\theta)$] & 0.997  & 1.001 $\pm$ 0.002  & 0.993  & 0.9975 $\pm$ 0.0005 \\
    ${\hat r_{\rm c}}$ [$r_{\rm c}/R_{\rm core}\,(\theta)$] & 1.013 & 1.005 $\pm$ 0.002   &  1.016  &  1.017 $\pm$ 0.002  \\
    ${\hat r_{\rm s}}$ [$r_{\rm s}/R_{\rm core}\,(\theta)$] & 1.013   & 1.011 $\pm$ 0.0005 & 1.016  & 1.0165 $\pm$ 0.0005 \\
    $\left(\rho_{\rm eq}/{}\rho_{\rm p}\right)_{\infty}$ & --   & 0.97 $\pm$ 0.07  & --  & 0.184 $\pm$ 0.001 \\
     \hline 
    \end{tabular}
    \end{center}
     \vspace{-0.6cm}
     \begin{flushleft} 
     $^a$ The fixed stellar parameters of the rotating central star for a given value of $V_{\rm rot}$ and $\theta$
     (i.e. $R_{\rm core}\,(\theta{})$,  $T_{\rm eff}\,(\theta{})$, $L\,(\theta{})$, $\Gamma\,(\theta{})$),
     resulting from its oblateness and the gravity darkening effect,
     are shown in the \emph{upper part} of the table.
     The converged wind parameters of each iteration process $v_{\infty}$, $\beta$, ${\hat g_{0}}$, $\gamma$, $\delta$, 
     ${\hat r}_{0}$, and mass-loss rates $\log\,\dot{M}$, together with the critical radius ${\hat r_{\rm c}}$
     and sonic radius ${\hat r_{\rm s}}$, the latter obtained by {\sc ISA-Wind},
     are displayed in the \emph{lower part} of the table. 
     The mean values for the equator in the second column were obtained from the parameter values
     of step no 11 and 18 of the one iteration cycle displayed in the lower part of Table~\ref{tablA2-wind-300kms},
     whereas the mean values for the equator in the fourth column, were obtained from the converged
     parameters of the two different iteration cycles displayed in the middle and lower part of Table~\ref{tablA3-wind-500kms},
     respectively.
     \end{flushleft}
    \end{table*}
\begin{figure*}
\centerline{\hbox{\hspace{0.2cm}\epsfxsize=8.7cm \epsfbox{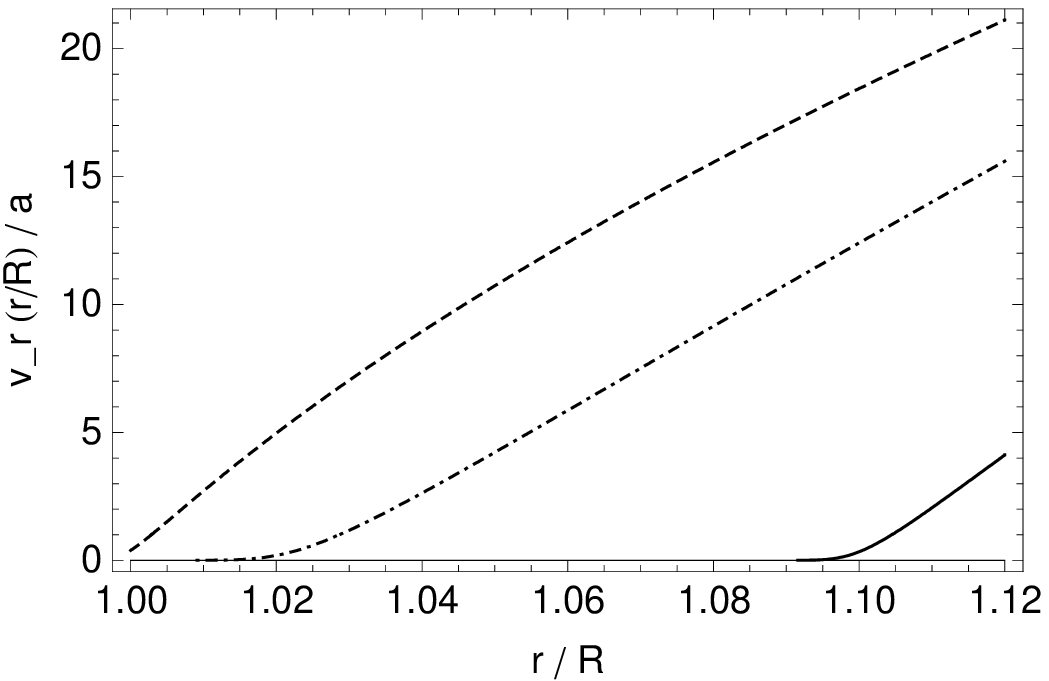}
\hspace{0.2cm} \epsfxsize=8.7cm \epsfbox{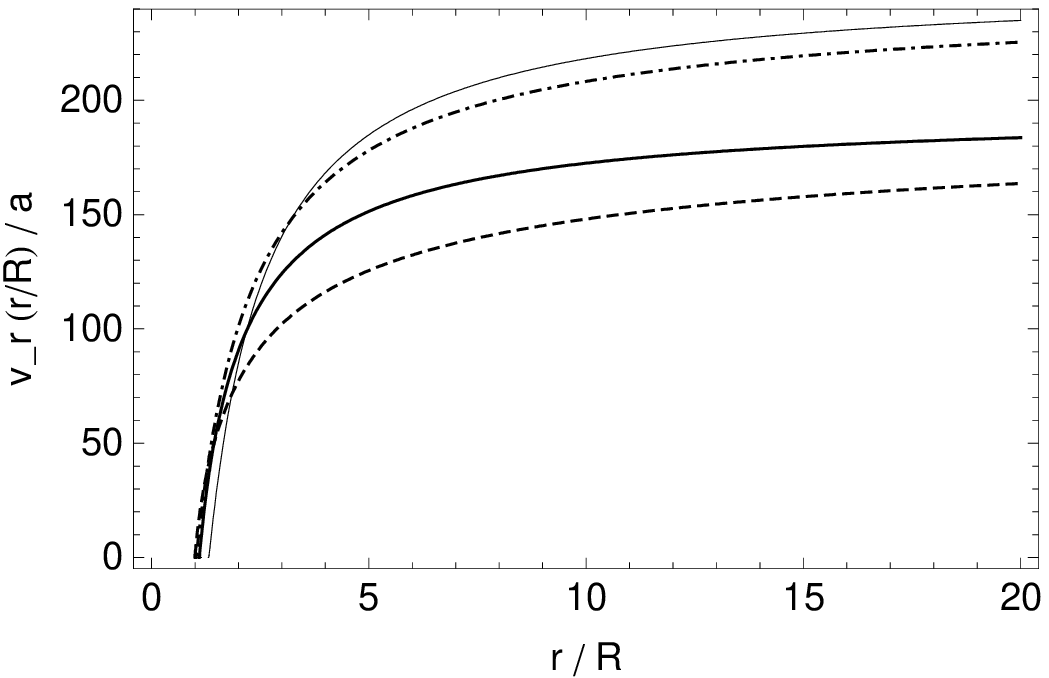}}}
\vspace{0.2cm}
\centerline{\hbox{\hspace{0cm}\epsfxsize=8.7cm \epsfbox{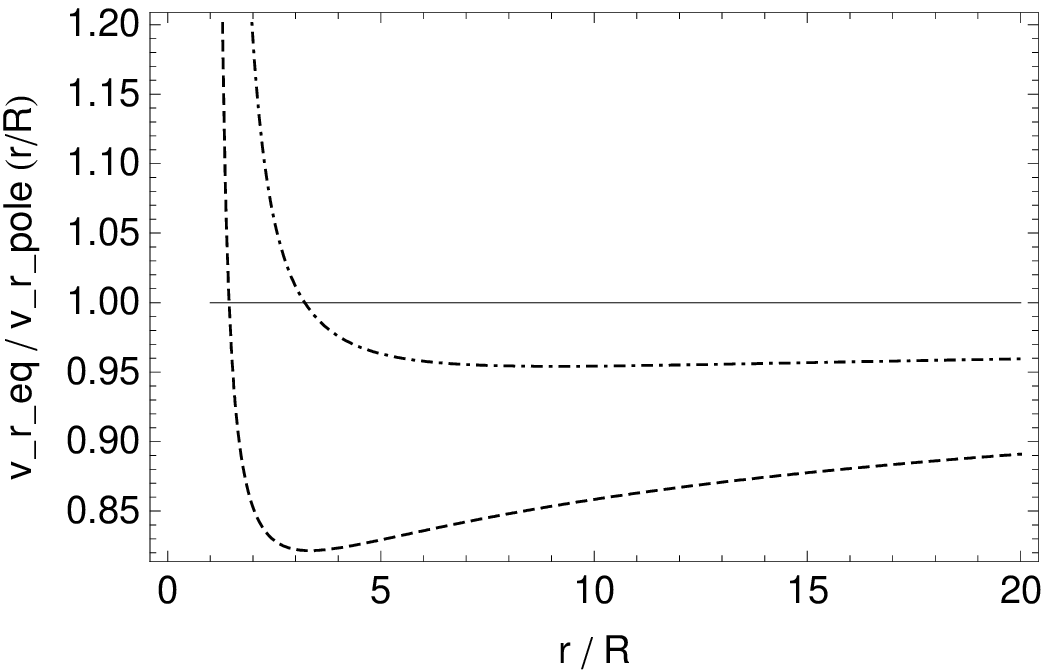}
\hspace{0.2cm} \epsfxsize=8.7cm \epsfbox{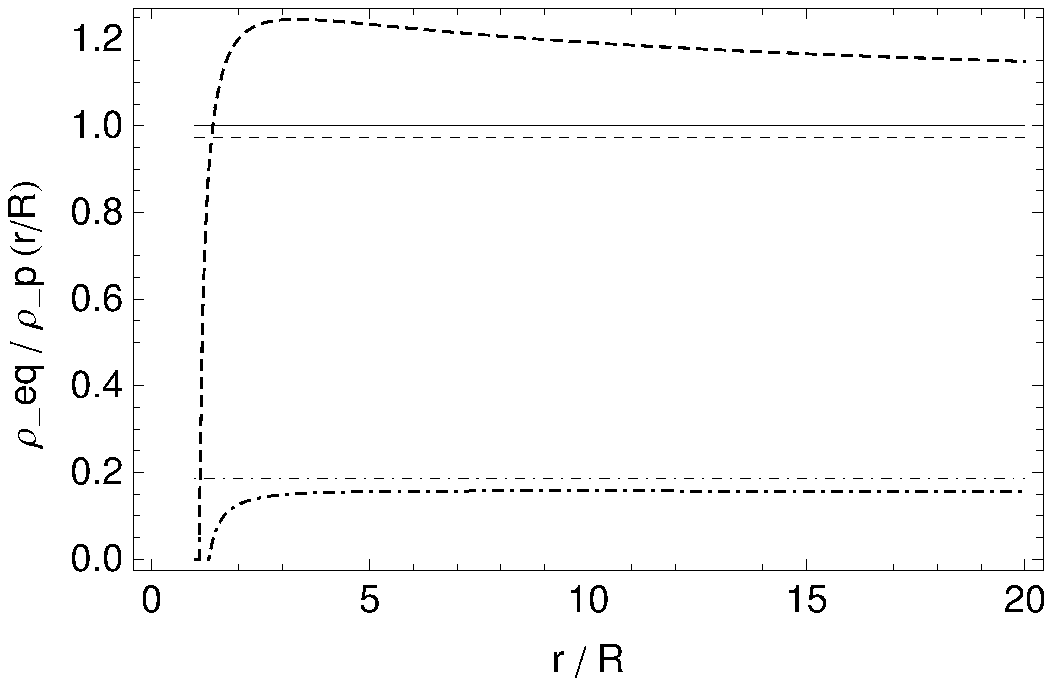}}}
\vfill\parbox[b]{18cm}{\caption[]{
      Model results for the wind from a differentially rotating O5--V main-sequence star 
      \emph{considering its oblateness including gravity darkening effects}
      with an equatorial rotation speed of $V_{\rm rot}=300$~km~s$^{-1}$ (see dashed curves) and 500~km~s$^{-1}$ (see dotted-dashed curves)
      \emph{at the equator} (for $\theta=\pi{}/2$),
      compared to the wind from the \emph{pole} at $\theta=0$ (see solid curves or solid horizontal lines, respectively);
      as for the stellar and wind parameters see Table~\ref{table2-par-dist-star} in Sect.~\ref{sect-applic-ostar}.
      All diagrams are plotted vs. radial distance ${\hat r}$ in units of the equatorial stellar core radius 
      $R_{\rm core}\,(V_{\rm rot},\theta{}=\pi/2)$.
      \emph{Upper panel}: the amount of the radial wind velocity component ${\hat v}_{\rm r}\,({\hat r},\theta{})$
      (in units of sound speed $a$) in the subsonic and lower supersonic wind regime (see \emph{left diagram}), and
      in the supersonic region up to stellar distances of ${\hat r}=20$ (see \emph{right diagram});
      the thick (thin) solid curves represent the radial wind velocity at the pole for $V_{\rm rot}=300$~km~s$^{-1}$ 
      (500~km~s$^{-1}$). 
      \emph{Lower panel}: the \emph{left diagram} displays the curves of the ratio 
      $v_{\rm r}\,({\hat r},\pi/2)/v_{\rm r}\,({\hat r},0)$ of the equatorial radial wind velocity compared to
      the polar wind velocity and the \emph{right diagram} shows the ratio $\rho\,({\hat r},\pi/2)/{}\rho\,({\hat r},0)$
      of the equatorial wind density in terms of the density at the pole.
      For the pole, these ratios are simply represented by the (solid) constant lines at $1.0$;
      at the equator the density ratios approach (for large stellar distances ${\hat r}$)
      the terminal values represented by the thin horizontal dashed or dotted-dashed line, respectively, in the right diagram.
      } \label{pic-vr-rho-dist-star} }
\end{figure*}

We now turn our attention to the physically more realistic cases, including the effects of stellar oblateness and gravity darkening. 
The model parameter and results are listed in Table~\ref{table2-par-dist-star} and the iteration cycles are shown 
in the Appendix. Again the results are also shown graphically (in Fig.~\ref{pic-vr-rho-dist-star}). 
The situation is notably different from the non-distorted case. 
Whilst the results for the 300 km/s ($\Omega = 0.70$) 
main-sequence dwarf are inconspicuous as they are not all that different from the spherical case, 
the behaviour of
$(\rho_{\rm eq}/{}\rho_{\rm p})$ again varies somewhat with stellar distance, having
its maximum of about 1.25 at about $r/R=$3.0, and then very slowly decreasing
to approach a value of 0.97 at infinity.
The more rapid rotator (at 500 km/s and $\Omega = 0.92$) shows, seemingly 
surprisingly, a lower mass-loss 
rate at the equator than at the pole, by almost 
an order of magnitude. 

The right bottom panel of Fig.~\ref{pic-vr-rho-dist-star} also shows the opposite effect: the ratio of
      equatorial-to-polar density ($\rho_{\rm eq}/\rho_{\rm p}$) is significantly below 1, by about a factor of 5. In other words, the 
predictions including gravity darkening suggest a polarward rather than equatorial stellar wind. How can we understand  
this result? As can be noted from Eq.~(\ref{local-gamma}) the Eddington parameter has become modified to correct for the 
latitudinal surface distortion. In comparison to the 1D spherical case the luminosity $L$ and effective temperature 
$T_{\rm eff}$ are lower
and using the results of Vink et al. (\cite{vink:al:00}) one can already expect the lowered equatorial mass-loss rate in comparison to the spherical case primarily due to the lower 
$L$. 
These results suggest that for rapid rotators, the big difference between the pole and equator would 
lead to significantly different diagnostics 
for a star that is observed from the pole or equator (see e.g. Hillier et al. \cite{hillier:lanz}, Herrero et al. \cite{herrero:garcia}). On a positive note, the 
pole-to-equator ratio of order 5 is sufficiently large to be measurable with linear polarisation data.

\section{Discussion and conclusions}   \label{sect-disc}
    
\subsection{Comparison with previous models}

The earliest works on CAK-type mass-loss predictions including the effects of stellar rotation via 
the effective mass in the equation of motion (but without considering gravity darkening) such 
as those by FA and Pauldrach et al. (\cite{pauldrach:al}), Petrenz \& Puls (\cite{petrenz:puls:2000}) 
resulted in moderate mass-loss rate enhancements, which were subsequently included in evolutionary calculations 
(Langer \cite{langer}). Our results, with mass-loss rates increases by at most 0.1 dex, are in good agreement with 
these earlier works based on the CAK theory. 

When gravity darkening is included, previous studies, such as those of Owocki et al. (\cite{owocki:cranmer}), Maeder (\cite{maeder}) and 
Pelupessy et al. (\cite{pelupessy:lamers:vink}) all found relatively modest changes for low rotational velocities, but they all favour 
polar ejection for high rotation rates in close proximity to the break-up limit ($\Omega$ = 1). 
The most-oft used formulae for mass-loss enhancement on the basis of the CAK theory is the one 
given by Maeder \& Meynet (\cite{maeder:meynet}). Their famous equation (4.29) reads:

\begin{equation}
\frac{\dot{M}(\Omega)}{\dot{M}(\Omega=0)} = \frac{(1-\Gamma)^{\frac{1}{\alpha}-1}}{\left[ 1-\frac{\Omega^2}{2\pi G \rho_{\rm m}} - \Gamma \right]^{\frac{1}{\alpha}-1}},
\label{EqMdotRot}
\end{equation}
where $\Gamma=L/L_{\rm Edd}=\kappa L / (4\pi cGM)$ is the Eddington factor (with $\kappa$ the total opacity), and $\alpha$ the $T_{\rm eff}-$dependent CAK 
force multiplier parameter.
Whilst it is claimed that the mass-loss increase is due to rotation ($\Omega$) the more relevant reason for the 
high mass-loss rate is in fact the high $\Gamma$ factor (see more recent computations by 
Vink \cite{vink}, Gr\"{a}fener \& Hamann \cite{graefener:hamann}, Vink et al. \cite{vink:muijres}).

Surprisingly, our 2D results that take gravity darkening into account predict a 
equatorial decrease in the mass-loss rate. This would also imply that the total (cf. Eq.~\ref{total-mass-loss}) 
mass-loss rate is now lower than for the spherical 1D case. 
This is in contradiction
to the above Maeder \& Meynet equation that is oftentimes 
used in stellar evolution calculations for rotating massive stars.

\subsection{Comparison with observations}

There are two complementary types of astronomical techniques that can 
provide meaningful geometric constraints on large-scale 2D wind structures. 
These are interferometry and linear spectropolarimetry. 
The current status is that whilst interferometry has been very successful in obtaining results 
on objects such as classical Be stars, B[e] supergiants, and LBVs, 
results for O stars are -- to the best of our knowledge --  not yet available due to their smaller radii. 

Linear spectropolarimetry can be employed with at least two levels of complexity. 
The first involves detailed line-profile predictions that can be used to 
determine the geometric and kinematic environments around young and massive stars 
(Harries \cite{harries}, Vink et al. \cite{vink:harries}), as well as the polarimetric agents (Kuhn et al. \cite{kuhn:berdyugina}). 
The lower-level complexity application of linear spectropolarimetry is adopted here, 
with the emission line reasonably assumed to be unpolarised, but with the 
continuum responsible for the observed polarisation level: depolarisation 
(Poeckert \& Marlborough \cite{poeckert:marlborough}, Brown \& McLean \cite{brown:mclean}). This latter approach is sufficient 
for our current comparisons. In this case a pole-to-equator density contrast of 5
might lead to a continuum polarisation of $\simeq$1\% according to the expectations of Harries et al. (\cite{harries:hillier}).

If we assume that our most realistic density-contrast predictions concern those models that 
include gravity darkening, we would anticipate the vast majority
of mostly moderately rotating O--dwarfs to be unpolarised, as the density contrast between the pole and equator 
is close to unity for rotation velocities up to 300 km/s. These overall findings are in good 
agreement with the linear H$\alpha$ polarimetry survey of Vink et al. (\cite{vink:davies}). 

An exception might involve the sub-group of Oe stars, for which Vink et al. (\cite{vink:davies}) list the highest rotation velocities, 
with $v$sin$i$ values approaching $\simeq$400 km/s. 
As this concerns a lower limit (due to the inclination effects), the true 
rotational velocities of Oe stars might go all the way up to the break-up limit ($\Omega = 1$). 
Interestingly, the Oe star HD\,45314 was indeed shown to have a line effect, but the overall incidence of line effects in Oe stars
was low: 1/6. 
On the basis of the general spectroscopic similarity between Oe and Be stars, and the presence of double-peaked line 
profiles many astronomers would expect Oe stars to be embedded in circumstellar disks, although it should be noted 
that spectroscopy alone cannot provide geometric constraints, 
without complementary imaging, interferometry, or polarimetry, because geometric 
information from spectral line profiles is not unique, due to the convolution that underlies spectral line formation.
In particular, double-peaked profiles can even be noted in spherical shells (e.g. PP).

In the current paper the surprising result is that of polar outflows, whilst it should be noted that 
linear spectropolarimetry cannot yet distinguish between a prolate 
(polar) or oblate (equatorial) wind structure (e.g. Wood et al. \cite{wood:bjorkman}).
It will be challenging to reconcile our axisymmetric density-contrast predictions with the general 
expectation that Oe/Be stars are embedded within equatorial structures. 
Clearly there are many puzzles remaining in the geometry and 
formation of circumstellar media around rotating OB stars, and quite possibly 
additional physical effects, such as magnetic fields, pulsations, and companions may need to be considered.

\subsection{Summary and Future Work}

In Paper I (M\"uller \& Vink \cite{mueller:vink}), we proposed a new parametrisation of the radiative line acceleration, 
expressing it as a function of radius rather than of the velocity gradient (as in CAK theory)
which generalised the classical thermal Parker (\cite{parker}) wind and allowed solving it analytically.
This has the advantage of higher accuracy and enables to check an influence of numerical viscosity if used as a test example for numerical simulations.
In addition, the implementation
of this formalism allows for local dynamical
consistency as we are able to determine the momentum
transfer at each location in the wind through the use of Monte
Carlo simulations. 

In this paper, we present a generalization of our model of spherical line-driven winds (in Paper I)
to the case of axially symmetric rotating stars.
We extend on the results of Paper~I,
deriving analytical solutions for the 2D case 
in an axisymmetric mass outflow or inflow scenario.
The generalization approximates the solution of the two-dimensional problem of a rotating wind 
as a one-parametric set (parameterised by the latitude) of one-dimensional solutions.
Here, the separation of the radial motion for individual latitudes is achieved by the basic assumption that the meridional flow velocity is negligible. 
Assuming furthermore only central external forces, then imply the conservation of angular momentum and leads to
an additional centrifugal term in the equation of radial motion.\footnote{ 
The centrifugal term was also included by Friend \& Castor (\cite{friend:castor}) in an opposite approximation of an extreme transport 
of angular momentum able to ensure purely radial motion in a co-rotating frame.} 
We also 
extend our iterative method for the simultaneous solution of the mass-loss rate and 
velocity field to the case of a rotating non-spherical stellar wind, using 
the parameterised description of the line acceleration that only depends 
on radius -- at any given latitude. 

Furthermore, an approximate analytical solution for the supersonic flow of a rotating wind 
is derived, which closely resembles the exact solution, and 
we apply the new expressions with our iterative method to (i) the stellar wind from a 
differentially rotating 40\,$M_{\sun}$ O5--V main sequence star, as well as to (ii) 
a 60\,$M_{\sun}$ O--giant star.
We subsequently account for the effects of stellar oblateness and gravity darkening for 
the O5--V main sequence star. The results show an 
equatorial decrease in the mass-loss rate, which 
would imply a total 
mass-loss rate that is lower than for the spherical 1D case, 
in contradiction to the oft-used Maeder \& Meynet (\cite{maeder:meynet}) formalism 
used in most current stellar evolution calculations for rotating massive stars. 

In the future we plan to extend our method to B supergiants, in particular 
to tackle the problem of disk-formation in B[e] supergiants (Lamers \& Pauldrach \cite{lamers:pauldrach}, Pelupessy et al. \cite{pelupessy:lamers:vink}, Cure
et al. \cite{cure:rial}, Madura et al. \cite{madura:owocki}).

    \begin{acknowledgements}
      We acknowledge financial report from the STFC and DCAL. We thank our anonymous referee for helpful comments.
    \end{acknowledgements}

    \appendix
     
     \section{Tables of iteration cycles}
          
%
    \begin{table*}
    \begin{center}
    \caption{Iteration cycles for the equatorial wind from a rotating O5--V main-sequence star
     with $V_{\rm rot}=300$~km s$^{-1}$ (see upper table) and $V_{\rm rot}=500$~km s$^{-1}$ (see lower table)
     without distortion.
     The variable stellar and wind parameters at each iteration step until convergence.$^a$}
    \label{tablA1-wind-non-dist}
     \begin{tabular}{rlllllllll}
   \hline\hline\noalign{\smallskip}
   Step &  $v_{\infty}$ & $\log\,\dot{M}$ & $v_{\infty ,\, \rm fit}$  & $\beta$ &  $\gamma_{\rm fit}$ & $\delta_{\rm fit}$ & ${\hat r}_{0, \rm fit}$
   & ${\hat r}_{0}'$ & ${\hat r}_{\rm s}$  \\
   no. & [km s$^{-1}$] &  [$M_{\sun}$/ yr] & [km s$^{-1}$] & & & & & & \\
    \hline\noalign{\smallskip} 
-1 & 2020  &    -5.500 &    --   &     1.0000 &     --     &     --     &     --     &     --     &      --    \\ 
 0 & 5805 &     -5.641 &   2365  &     0.8664 &     0.7329 &     0.4917 &     1.0008 &     1.0066 &     1.0175 \\
 1 & 4063 &     -6.154 &   3633  &     0.7929 &     0.5859 &     0.7077 &     1.0026 &     1.0086 &     1.0102 \\
 2 & 3002 &     -6.262 &   4077  &     0.7440 &     0.4879 &     0.7669 &     1.0039 &     1.0098 &     1.0112 \\
 3 & 2554 &     -6.199 &   3818  &     0.7279 &     0.4557 &     0.7563 &     1.0020 &     1.0101 &     1.0114 \\
 4 & 2393 &     -6.112 &   3436  &     0.7347 &     0.4693 &     0.7402 &     0.9985 &     1.0099 &     1.0114 \\
 5 & 2476 &     -6.047 &   3175  &     0.7486 &     0.4972 &     0.7262 &     0.9967 &     1.0097 &     1.0113 \\
 6 & 2843 &     -6.026 &   3143  &     0.7385 &     0.4769 &     0.6859 &     1.0012 &     1.0095 &     1.0113 \\
 7 & 2666 &     -6.064 &   3287  &     0.7310 &     0.4619 &     0.6992 &     1.0006 &     1.0096 &     1.0109 \\
 8 & 2419 &     -6.061 &   3215  &     0.7461 &     0.4923 &     0.7376 &     0.9965 &     1.0097 &     1.0110 \\
 9 & 2542 &     -6.026 &   3032  &     0.7696 &     0.5393 &     0.7554 &     0.9953 &     1.0095 &     1.0113 \\
10 & 2915 &     -6.026 &   3086  &     0.7574 &     0.5149 &     0.7156 &     1.0005 &     1.0093 &     1.0114 \\
     \hline 
    \end{tabular}
    
    \vspace{1cm}
    
    \begin{tabular}{rlllllllll}
   \hline\hline\noalign{\smallskip}
   Step &  $v_{\infty}$ & $\log\,\dot{M}$ & $v_{\infty ,\, \rm fit}$  & $\beta$ &  $\gamma_{\rm fit}$ & $\delta_{\rm fit}$ & ${\hat r}_{0, \rm fit}$
   & ${\hat r}_{0}'$ & ${\hat r}_{\rm s}$  \\
   no. & [km s$^{-1}$] &  [$M_{\sun}$/ yr] & [km s$^{-1}$] & & & & & & \\
    \hline\noalign{\smallskip} 
-1 & 2020 &     -5.500 &   --   &     1.0000 &     --     &     --     &     --     &     --     &      --    \\ 
 0 & 4553 &     -5.641 &   2365 &     0.8664 &     0.7329 &     0.4917 &     1.0008 &     1.0066 &     1.0175 \\
 1 & 2930 &     -6.057 &   3148 &     0.8132 &     0.6265 &     0.7377 &     1.0005 &     1.0084 &     1.0105 \\
 2 & 2513 &     -6.095 &   3303 &     0.7799 &     0.5598 &     0.7572 &     1.0022 &     1.0093 &     1.0117 \\
 3 & 2116 &     -6.059 &   3146 &     0.7751 &     0.5501 &     0.7580 &     0.9984 &     1.0094 &     1.0117 \\
 4 & 2186 &     -5.982 &   2860 &     0.7727 &     0.5454 &     0.7265 &     0.9987 &     1.0092 &     1.0119 \\
 5 & 2324 &     -5.953 &   2822 &     0.7652 &     0.5304 &     0.6898 &     1.0000 &     1.0091 &     1.0117 \\
 6 & 1996 &     -5.957 &   2787 &     0.7859 &     0.5717 &     0.7361 &     0.9932 &     1.0091 &     1.0114 \\
 7 & 2262 &     -5.907 &   2608 &     0.7949 &     0.5898 &     0.7171 &     0.9962 &     1.0089 &     1.0121 \\
 8 & 2248 &     -5.923 &   2658 &     0.8031 &     0.6063 &     0.7337 &     0.9950 &     1.0089 &     1.0117 \\
 9 & 2372 &     -5.929 &   2706 &     0.7951 &     0.5903 &     0.7144 &     0.9975 &     1.0088 &     1.0119 \\
10 & 2227 &     -5.954 &   2767 &     0.7934 &     0.5868 &     0.7292 &     0.9960 &     1.0089 &     1.0116 \\
11 & 2176 &     -5.944 &   2707 &     0.8085 &     0.6170 &     0.7513 &     0.9936 &     1.0089 &     1.0118 \\
12 & 2334 &     -5.930 &   2652 &     0.8106 &     0.6213 &     0.7439 &     0.9959 &     1.0088 &     1.0121 \\
13 & 2317 &     -5.948 &   2728 &     0.8189 &     0.6378 &     0.7493 &     0.9943 &     1.0088 &     1.0119 \\
14 & 2416 &     -5.956 &   2763 &     0.8109 &     0.6217 &     0.7408 &     0.9969 &     1.0088 &     1.0121 \\
15 & 2528 &     -5.974 &   2887 &     0.7818 &     0.5637 &     0.7017 &     1.0006 &     1.0089 &     1.0119 \\
16 & 2183 &     -5.997 &   2928 &     0.7853 &     0.5706 &     0.7507 &     0.9973 &     1.0092 &     1.0114 \\
17 & 2058 &     -5.962 &   2744 &     0.8115 &     0.6230 &     0.7657 &     0.9914 &     1.0091 &     1.0119 \\
18 & 2299 &     -5.922 &   2606 &     0.8278 &     0.6557 &     0.7611 &     0.9934 &     1.0088 &     1.0124 \\
19 & 2613 &     -5.937 &   2720 &     0.8081 &     0.6163 &     0.7274 &     0.9993 &     1.0087 &     1.0123 \\
     \hline 
    \end{tabular}
    \end{center}
     \begin{flushleft} 
     $^a$ For the fixed stellar parameters $L$, $T_{\rm eff}$, $R$, $M$, and $\Gamma$, see upper part of Table~\ref{table1-par-nodist-star} in
     Sect.~\ref{sect-applic-ostar}.
     The line acceleration parameters $\gamma_{\rm fit}$ (or equivalently $\beta$),
     $\delta_{\rm fit}$ and ${\hat r_{0, \rm fit}}$, and the terminal velocity $v_{\infty ,\,\rm fit}$,
     were determined by (a simplified version of) fitting formula Eq.~(\ref{line-acc-term3a}),
     applied to the results from a numerical calculation of the line acceleration ${\hat g_{\rm rad}^{\rm line}}\,({\hat r}_{i},\theta{})$
     at the equator ($\theta{}=\pi{}/2$).
     The parameter ${\hat r}_{0}'$ (in the $\beta$ velocity law, Eq.~\ref{beta-law}) and the sonic radius ${\hat r}_{\rm s}$
     are output values from {\sc ISA-Wind}, whereas $\log\,\dot{M}$ is the improved estimated mass-loss rate numerically obtained
     (by {\sc MC-Wind} and Eq.~\ref{Mdot-numerical}).
     At each iteration step, the value of $v_{\infty}$ was calculated by (a simplified version of) Eq.~(\ref{v-inf-next})
     and used as the new input value for the terminal velocity in the next iteration step,
     together with the new estimates of $\dot{M}$ and $\beta$ (cf. description of iteration process in Sect.~\ref{sect-it-method}).
     Convergence is achieved when 
     the values of $v_{\infty ,\,\rm fit}$ and $v_{\infty}$ have become equal or close to each other.
     In this case, the condition that the critical radius ${\hat r}_{\rm c}$ (determined by Eq.~\ref{crit-rad-eq})
     has to be equal to the sonic radius ${\hat r}_{\rm s}$ (determined by {\sc ISA-Wind})
     is fulfilled.
    \end{flushleft}
    \end{table*}
%
%

%
    \begin{table*}
    \begin{center}
    \caption{Iteration cycles for the wind from an O--giant star rotating
     with $V_{\rm rot}=300$~km s$^{-1}$ at the \emph{pole} (see \emph{upper} table) and at the equator (see lower table)
     without distortion.
     The variable stellar and wind parameters at each iteration step until convergence.$^a$}
    \label{tabl-A1b-giant-wind}
     \begin{tabular}{rlllllllll}
   \hline\hline\noalign{\smallskip}
   Step &  $v_{\infty}$ & $\log\,\dot{M}$ & $v_{\infty ,\, \rm fit}$  & $\beta$ &  $\gamma_{\rm fit}$ & $\delta_{\rm fit}$ & ${\hat r}_{0, \rm fit}$
   & ${\hat r}_{0}'$ & ${\hat r}_{\rm s}$  \\
   no. & [km s$^{-1}$] &  [$M_{\sun}$/ yr] & [km s$^{-1}$] & & & & & & \\
    \hline\noalign{\smallskip} 
-1 & 3000 &     -4.700 &   --   &     1.0000 &       --   &       --   &       --   &      --    &      --    \\    
 0 & 5137 &     -4.998 &   2622 &     0.8734 &     0.7468 &     0.5186 &     1.0023 &     1.0104 &     1.0180 \\
 1 & 3087 &     -5.427 &   3504 &     0.8442 &     0.6885 &     0.7625 &     1.0001 &     1.0134 &     1.0153 \\
 2 & 2680 &     -5.429 &   3525 &     0.8546 &     0.7092 &     0.8491 &     0.9984 &     1.0145 &     1.0173 \\
 3 & 2806 &     -5.377 &   3306 &     0.8685 &     0.7370 &     0.8434 &     0.9983 &     1.0143 &     1.0178 \\
 4 & 3084 &     -5.369 &   3311 &     0.8534 &     0.7067 &     0.8049 &     1.0024 &     1.0140 &     1.0178 \\
 5 & 2843 &     -5.402 &   3399 &     0.8564 &     0.7129 &     0.8350 &     0.9999 &     1.0142 &     1.0172 \\
 6 & 3010 &     -5.388 &   3371 &     0.8456 &     0.6912 &     0.8117 &     1.0028 &     1.0142 &     1.0176 \\
 7 & 2654 &     -5.401 &   3379 &     0.8700 &     0.7400 &     0.8628 &     0.9960 &     1.0143 &     1.0173 \\
 8 & 3081 &     -5.361 &   3280 &     0.8525 &     0.7051 &     0.7982 &     1.0024 &     1.0140 &     1.0180 \\
     \hline 
    \end{tabular}
    
    \vspace{1cm}
    
      \begin{tabular}{rlllllllll}
   \hline\hline\noalign{\smallskip}
   Step &  $v_{\infty}$ & $\log\,\dot{M}$ & $v_{\infty ,\, \rm fit}$  & $\beta$ &  $\gamma_{\rm fit}$ & $\delta_{\rm fit}$ & ${\hat r}_{0, \rm fit}$
   & ${\hat r}_{0}'$ & ${\hat r}_{\rm s}$  \\
   no. & [km s$^{-1}$] &  [$M_{\sun}$/ yr] & [km s$^{-1}$] & & & & & & \\
    \hline\noalign{\smallskip} 
-1 & 3000 &     -4.700 &    --   &     1.0000 &       --   &       --   &       --   &      --    &      --    \\    
 0 & 4284 &     -4.998 &   2622 &     0.8751 &     0.7503 &     0.5198 &     1.0022 &     1.0104 &     1.0180 \\
 1 & 2674 &     -5.332 &   3212 &     0.8433 &     0.6866 &     0.7494 &     1.0013 &     1.0132 &     1.0156 \\
 2 & 2304 &     -5.326 &   3144 &     0.8509 &     0.7018 &     0.8030 &     0.9986 &     1.0140 &     1.0173 \\
 3 & 2384 &     -5.262 &   2951 &     0.8508 &     0.7017 &     0.7648 &     0.9987 &     1.0137 &     1.0179 \\
 4 & 2430 &     -5.229 &   2882 &     0.8484 &     0.6968 &     0.7557 &     0.9992 &     1.0136 &     1.0176 \\
 5 & 2447 &     -5.203 &   2911 &     0.8434 &     0.6868 &     0.7232 &     0.9986 &     1.0135 &     1.0174 \\
 6 & 2576 &     -5.185 &   2955 &     0.8235 &     0.6470 &     0.6754 &     1.0009 &     1.0135 &     1.0172 \\
 7 & 2422 &     -5.198 &   2973 &     0.8136 &     0.6273 &     0.6819 &     1.0003 &     1.0136 &     1.0167 \\
 8 & 2329 &     -5.177 &   2912 &     0.8125 &     0.6250 &     0.6818 &     0.9992 &     1.0137 &     1.0168 \\
 9 & 2505 &     -5.144 &   2945 &     0.7939 &     0.5877 &     0.6153 &     1.0012 &     1.0136 &     1.0168 \\
10 & 2358 &     -5.154 &   2997 &     0.7832 &     0.5663 &     0.6105 &     1.0003 &     1.0137 &     1.0163 \\
11 & 2306 &     -5.134 &   2928 &     0.7765 &     0.5530 &     0.6071 &     1.0004 &     1.0138 &     1.0164 \\
12 & 2532 &     -5.113 &   3021 &     0.7488 &     0.4977 &     0.5316 &     1.0030 &     1.0137 &     1.0162 \\
13 & 1994 &     -5.140 &   2964 &     0.7599 &     0.5197 &     0.5962 &     0.9968 &     1.0140 &     1.0157 \\
14 & 2369 &     -5.062 &   2905 &     0.7431 &     0.4863 &     0.4975 &     1.0007 &     1.0137 &     1.0164 \\
15 & 2366 &     -5.083 &   2975 &     0.7290 &     0.4580 &     0.5025 &     1.0023 &     1.0138 &     1.0157 \\
16 & 2274 &     -5.093 &   3021 &     0.7225 &     0.4450 &     0.5036 &     1.0021 &     1.0140 &     1.0156 \\
17 & 2235 &     -5.082 &   3000 &     0.7167 &     0.4335 &     0.4935 &     1.0020 &     1.0140 &     1.0157 \\
18 & 2190 &     -5.069 &   2968 &     0.7101 &     0.4202 &     0.4827 &     1.0018 &     1.0140 &     1.0156 \\
19 & 2213 &     -5.055 &   2976 &     0.7030 &     0.4060 &     0.4598 &     1.0019 &     1.0140 &     1.0156 \\
20 & 2059 &     -5.051 &   2941 &     0.7086 &     0.4172 &     0.4690 &     0.9995 &     1.0140 &     1.0155 \\
21 & 2141 &     -5.023 &   2920 &     0.7090 &     0.4180 &     0.4396 &     0.9991 &     1.0139 &     1.0156 \\
22 & 1988 &     -5.022 &   2823 &     0.7304 &     0.4607 &     0.4808 &     0.9949 &     1.0138 &     1.0154 \\
23 & 2453 &     -4.996 &   2855 &     0.7137 &     0.4275 &     0.4261 &     1.0015 &     1.0136 &     1.0157 \\
     \hline 
    \end{tabular}   
    \end{center}
     \begin{flushleft} 
     $^a$ For the fixed stellar parameters $L\,(\theta)$, $T_{\rm eff}\,(\theta)$, $R\,(\theta)$, $M$, and $\Gamma\,(\theta)$
     at the pole and the equator, respectively, see upper part of Table~\ref{table1b-par-nodist-star} in 
     Sect.~\ref{sect-applic-ostar}.
     The line acceleration parameters $\gamma_{\rm fit}$ (or equivalently $\beta$),
     $\delta_{\rm fit}$ and ${\hat r_{0, \rm fit}}$, and the terminal velocity $v_{\infty ,\,\rm fit}$,
     were determined by (a simplified version of) fitting formula Eq.~(\ref{line-acc-term3a}),
     applied to the results from a numerical calculation of the line acceleration ${\hat g_{\rm rad}^{\rm line}}\,({\hat r}_{i},\theta{})$
     at the pole ($\theta{}=0$) or at the equator ($\theta{}=\pi{}/2$), respectively.
     The parameter ${\hat r}_{0}'$ (in the $\beta$ velocity law, Eq.~\ref{beta-law}) and the sonic radius ${\hat r}_{\rm s}$
     are output values from {\sc ISA-Wind}, whereas $\log\,\dot{M}$ is the improved estimated mass-loss rate numerically obtained
     (by {\sc MC-Wind} and Eq.~\ref{Mdot-numerical}); 
     see also description of the converging iteration process in Table~\ref{tablA1-wind-non-dist}.
    \end{flushleft}
    \end{table*}
%
%


%
    \begin{table*}
    \begin{center}
    \caption{Iteration cycles for the wind from a rotating O5--V main-sequence star
     with $V_{\rm rot}=300$~km s$^{-1}$ at the \emph{pole} (see upper table) and at the equator (see lower table)
     considering the stellar distortion and the effects of gravity darkening.
     The variable stellar and wind parameters at each iteration step until convergence.$^a$}
    \label{tablA2-wind-300kms}
     \begin{tabular}{rlllllllll}
   \hline\hline\noalign{\smallskip}
   Step &  $v_{\infty}$ & $\log\,\dot{M}$ & $v_{\infty ,\, \rm fit}$  & $\beta$ &  $\gamma_{\rm fit}$ & $\delta_{\rm fit}$ & ${\hat r}_{0, \rm fit}$
   & ${\hat r}_{0}'$ & ${\hat r}_{\rm s}$  \\
   no. & [km s$^{-1}$] &  [$M_{\sun}$/ yr] & [km s$^{-1}$] & & & & & & \\
    \hline\noalign{\smallskip} 
-1 & 2020  &     -5.500 &    --   &     1.0000 &     --     &     --     &     --     &     --     &      --    \\ 
 0 & 5815  &     -5.583 &   2298  &     0.8836 &     0.7672 &     0.5874 &     1.0009 &     1.0072 &     1.0182 \\
 1 & 4363  &     -6.078 &   3707  &     0.8154 &     0.6308 &     0.7506 &     1.0015 &     1.0090 &     1.0109 \\
 2 & 3164  &     -6.202 &   4231  &     0.8061 &     0.6123 &     0.8630 &     0.9978 &     1.0103 &     1.0118 \\
 3 & 2955  &     -6.137 &   3856  &     0.8277 &     0.6554 &     0.9134 &     0.9937 &     1.0104 &     1.0125 \\
 4 & 3371  &     -6.078 &   3633  &     0.8280 &     0.6559 &     0.8794 &     0.9982 &     1.0100 &     1.0127 \\
 5 & 3390  &     -6.098 &   3741  &     0.8181 &     0.6361 &     0.8646 &     0.9990 &     1.0100 &     1.0123 \\
 6 & 3080  &     -6.110 &   3725  &     0.8415 &     0.6830 &     0.9181 &     0.9930 &     1.0101 &     1.0122 \\
 7 & 3590  &     -6.080 &   3621  &     0.8345 &     0.6690 &     0.8870 &     0.9997 &     1.0099 &     1.0127 \\
 8 & 3493  &     -6.124 &   3857  &     0.8159 &     0.6318 &     0.8596 &     1.0001 &     1.0100 &     1.0123 \\
 9 & 3214  &     -6.136 &   3874  &     0.8191 &     0.6383 &     0.8873 &     0.9976 &     1.0102 &     1.0122 \\
10 & 3251  &     -6.109 &   3785  &     0.8175 &     0.6350 &     0.8685 &     0.9977 &     1.0102 &     1.0124 \\
11 & 3023  &     -6.099 &   3660  &     0.8431 &     0.6862 &     0.9255 &     0.9922 &     1.0101 &     1.0123 \\
12 & 3263  &     -6.072 &   3502  &     0.8602 &     0.7205 &     0.9264 &     0.9933 &     1.0099 &     1.0128 \\
13 & 3592  &     -6.087 &   3607  &     0.8534 &     0.7068 &     0.8978 &     0.9974 &     1.0098 &     1.0128 \\
     \hline 
    \end{tabular}
    
    \vspace{1cm}
    
      \begin{tabular}{rlllllllll}
   \hline\hline\noalign{\smallskip}
   Step &  $v_{\infty}$ & $\log\,\dot{M}$ & $v_{\infty ,\, \rm fit}$  & $\beta$ &  $\gamma_{\rm fit}$ & $\delta_{\rm fit}$ & ${\hat r}_{0, \rm fit}$
   & ${\hat r}_{0}'$ & ${\hat r}_{\rm s}$  \\
   no. & [km s$^{-1}$] &  [$M_{\sun}$/ yr] & [km s$^{-1}$] & & & & & & \\
    \hline\noalign{\smallskip} 
-1 & 4000  &  -6.000  &   --    &    1.1000 &     --      &    --      &    --      &    --      &      --    \\ 
 0 & 6210  &  -6.241  &   3761  &    0.9279 &     0.8558  &    0.6448  &    1.0022  &    1.0077  &    1.0161 \\
 1 & 5992  &  -6.509  &   6284  &    0.7568 &     0.5137  &    0.4524  &    1.0042  &    1.0096  &    1.0118 \\ 
 2 & 6342  &  -6.624  &   6825  &    0.6393 &     0.2787  &    0.4487  &    1.0051  &    1.0110  &    1.0110 \\
 3 & 1495  &  -6.704  &   5275  &    0.6231 &     0.2463  &    0.9999  &    1.0063  &    1.0115  &    1.0116 \\
 4 & 2085  &  -6.277  &   3789  &    0.6321 &     0.2642  &    0.5015  &    1.0024  &    1.0122  &    1.0123 \\
 5 & 2212  &  -6.122  &   3641  &    0.6521 &     0.3042  &    0.4549  &    0.9980  &    1.0105  &    1.0111 \\
 6 & 2422  &  -6.058  &   3451  &    0.6795 &     0.3590  &    0.4598  &    0.9965  &    1.0101  &    1.0109 \\
 7 & 2566  &  -6.057  &   3398  &    0.6984 &     0.3967  &    0.4858  &    0.9968  &    1.0099  &    1.0108 \\
 8 & 2776  &  -6.076  &   3544  &    0.6981 &     0.3962  &    0.4722  &    0.9987  &    1.0098  &    1.0108 \\
 9 & 2795  &  -6.116  &   3738  &    0.6854 &     0.3708  &    0.4600  &    1.0001  &    1.0099  &    1.0108 \\
10 & 2500  &  -6.140  &   3752  &    0.6829 &     0.3658  &    0.4824  &    0.9983  &    1.0101  &    1.0109 \\
11 & 3177  &  -6.111  &   3883  &    0.6557 &     0.3114  &    0.4021  &    1.0029  &    1.0101  &    1.0110 \\
12 & 2303 &   -6.186 &   4011 &     0.6543 &     0.3086 &     0.4602 &     0.9991 &     1.0102 &     1.0107 \\
13 & 2718 &   -6.105 &   3839 &     0.6464 &     0.2928 &     0.4014 &     1.0017 &     1.0102 &     1.0110 \\
14 & 2380 &   -6.123 &   3823 &     0.6509 &     0.3018 &     0.4323 &     0.9993 &     1.0102 &     1.0107 \\
15 & 2396 &   -6.082 &   3584 &     0.6720 &     0.3440 &     0.4563 &     0.9972 &     1.0101 &     1.0108 \\
16 & 2671 &   -6.064 &   3533 &     0.6791 &     0.3581 &     0.4500 &     0.9994 &     1.0100 &     1.0109 \\
17 & 2422 &   -6.094 &   3576 &     0.6986 &     0.3973 &     0.4875 &     0.9943 &     1.0100 &     1.0108 \\
18 & 2833 &   -6.077 &   3537 &     0.6978 &     0.3956 &     0.4669 &     0.9992 &     1.0099 &     1.0110 \\
     \hline 
    \end{tabular}   
    \end{center}
     \begin{flushleft} 
     $^a$ For the fixed stellar parameters $L\,(\theta)$, $T_{\rm eff}\,(\theta)$, $R\,(\theta)$, $M$, and $\Gamma\,(\theta)$
     at the pole and the equator, respectively, see upper part of Table~\ref{table2-par-dist-star} (the first two columns) in 
     Sect.~\ref{sect-applic-ostar}.
     The line acceleration parameters $\gamma_{\rm fit}$ (or equivalently $\beta$),
     $\delta_{\rm fit}$ and ${\hat r_{0, \rm fit}}$, and the terminal velocity $v_{\infty ,\,\rm fit}$,
     were determined by (a simplified version of) fitting formula Eq.~(\ref{line-acc-term3a}),
     applied to the results from a numerical calculation of the line acceleration ${\hat g_{\rm rad}^{\rm line}}\,({\hat r}_{i},\theta{})$
     at the pole ($\theta{}=0$) or at the equator ($\theta{}=\pi{}/2$), respectively.
     The parameter ${\hat r}_{0}'$ (in the $\beta$ velocity law, Eq.~\ref{beta-law}) and the sonic radius ${\hat r}_{\rm s}$
     are output values from {\sc ISA-Wind}, whereas $\log\,\dot{M}$ is the improved estimated mass-loss rate numerically obtained
     (by {\sc MC-Wind} and Eq.~\ref{Mdot-numerical}); 
     see also description of the converging iteration process in Table~\ref{tablA1-wind-non-dist}.
    \end{flushleft}
    \end{table*}
%
%

%
    \begin{table*}
    \begin{center}
    \caption{Iteration cycles for the wind from a rotating O5--V main-sequence star
     with $V_{\rm rot}=500$~km s$^{-1}$ at the \emph{pole} (see upper table) and at the equator (see middle and lower table)
     considering the stellar distortion and the effects of gravity darkening.
     The variable stellar and wind parameters at each iteration step until convergence.$^a$}
    \label{tablA3-wind-500kms}
     \begin{tabular}{rlllllllll}
   \hline\hline\noalign{\smallskip}
   Step &  $v_{\infty}$ & $\log\,\dot{M}$ & $v_{\infty ,\, \rm fit}$  & $\beta$ &  $\gamma_{\rm fit}$ & $\delta_{\rm fit}$ & ${\hat r}_{0, \rm fit}$
   & ${\hat r}_{0}'$ & ${\hat r}_{\rm s}$  \\
   no. & [km s$^{-1}$] &  [$M_{\sun}$/ yr] & [km s$^{-1}$] & & & & & & \\
    \hline\noalign{\smallskip} 
-1 & 2020  &     -5.500 &    --   &     1.0000 &     --     &     --     &     --     &     --     &      --    \\ 
 0 & 4952  &     -5.522 &   2243  &     1.0141 &     1.0282 &     0.8806 &     0.9917 &     1.0086 &     1.0199 \\
 1 & 4994  &     -5.875 &   3566  &     0.9915 &     0.9829 &     0.9326 &     0.9928 &     1.0095 &     1.0144 \\
 2 & 4519  &     -6.049 &   4515  &     0.9739 &     0.9478 &     0.9574 &     0.9926 &     1.0106 &     1.0149 \\
 3 & 4247  &     -6.091 &   4779  &     0.9916 &     0.9832 &     0.9929 &     0.9885 &     1.0110 &     1.0153 \\
 4 & 4311  &     -6.087 &   4695  &     1.0149 &     1.0297 &     1.0280 &     0.9874 &     1.0109 &     1.0160 \\
 5 & 4715  &     -6.092 &   4755  &     0.9999 &     0.9999 &     0.9975 &     0.9935 &     1.0108 &     1.0164 \\
     \hline 
    \end{tabular}
    
    \vspace{1cm}
    
\begin{tabular}{rlllllllll}
   \hline\hline\noalign{\smallskip}
   Step &  $v_{\infty}$ & $\log\,\dot{M}$ & $v_{\infty ,\, \rm fit}$  & $\beta$ &  $\gamma_{\rm fit}$ & $\delta_{\rm fit}$ & ${\hat r}_{0, \rm fit}$
   & ${\hat r}_{0}'$ & ${\hat r}_{\rm s}$  \\
   no. & [km s$^{-1}$] &  [$M_{\sun}$/ yr] & [km s$^{-1}$] & & & & & & \\
    \hline\noalign{\smallskip} 
-1 & 2020  &     -5.500 &   --   &     1.0000 &     --     &     --     &     --     &     --     &      --    \\ 
 0 & 2558  &     -6.055 &   1176 &     0.9542 &     0.9084 &     0.5865 &     0.9823 &     1.0054 &     1.0147 \\
 1 & 2146  &     -6.452 &   1824 &     1.1567 &     1.3134 &     0.8351 &     0.9410 &     1.0068 &     1.0125 \\
 2 & 3199  &     -6.614 &   2199 &     1.1669 &     1.3338 &     0.7990 &     0.9747 &     1.0062 &     1.0227 \\
 3 & 3616  &     -6.813 &   2989 &     1.1168 &     1.2337 &     0.8633 &     0.9896 &     1.0068 &     1.0189 \\
 4 & 3791  &     -6.916 &   3847 &     1.0401 &     1.0802 &     0.8236 &     0.9973 &     1.0075 &     1.0164 \\
     \hline 
    \end{tabular}    
    
    \vspace{1cm}
    
     \begin{tabular}{rlllllllll}
   \hline\hline\noalign{\smallskip}
   Step &  $v_{\infty}$ & $\log\,\dot{M}$ & $v_{\infty ,\, \rm fit}$  & $\beta$ &  $\gamma_{\rm fit}$ & $\delta_{\rm fit}$ & ${\hat r}_{0, \rm fit}$
   & ${\hat r}_{0}'$ & ${\hat r}_{\rm s}$  \\
   no. & [km s$^{-1}$] &  [$M_{\sun}$/ yr] & [km s$^{-1}$] & & & & & & \\
    \hline\noalign{\smallskip} 
-1 & 2020 &     -5.500 &   --   &     1.0000 &     --     &     --     &     --     &     --     &      --    \\ 
 0 & 2594 &     -6.055 &   1195 &     0.9451 &     0.8903 &     0.5727 &     0.9838 &     1.0054 &     1.0147 \\
 1 & 2172 &     -6.455 &   1836 &     1.1381 &     1.2761 &     0.8318 &     0.9462 &     1.0069 &     1.0122 \\
 2 & 2946 &     -6.619 &   2150 &     1.2069 &     1.4138 &     0.8533 &     0.9642 &     1.0064 &     1.0216 \\
 3 & 3718 &     -6.798 &   2821 &     1.1447 &     1.2894 &     0.8726 &     0.9894 &     1.0065 &     1.0215 \\
 4 & 4036 &     -6.920 &   3828 &     1.0562 &     1.1123 &     0.8364 &     0.9983 &     1.0073 &     1.0170 \\
     \hline 
    \end{tabular}
    \end{center}    
   \begin{flushleft} 
    $^a$ The different results in the middle and lower table for the same latitude (i.e. equator)
     were obtained by different values for the number of photon packets ($N_{\rm ph}=2.0\times{}10^{7}$ in the middle
     and $N_{\rm ph}=3.5\times{}10^{7}$ in the lower table, respectively).
     For the fixed stellar parameters $L\,(\theta)$, $T_{\rm eff}\,(\theta)$, $R\,(\theta)$, $M$, and $\Gamma\,(\theta)$
     at the pole and the equator, respectively, see upper part of Table~\ref{table2-par-dist-star} (the last two columns) in 
     Sect.~\ref{sect-applic-ostar};
     see also analogous description of parameters in Table~\ref{tablA1-wind-non-dist}. 
   \end{flushleft}
    \end{table*}

\end{document}